\documentclass[preprint,12pt]{elsarticle}

\usepackage[top=1in, bottom=1in, left=1in, right=1in]{geometry}

\usepackage{graphicx}
\usepackage{amsmath}

\usepackage{natbib}
\usepackage{amssymb}

\usepackage{amsthm}

\usepackage{lineno}
\usepackage{subfig}
\usepackage{enumerate}
\usepackage{fullpage}
\usepackage{algorithm}
\usepackage{algpseudocode}
\usepackage[colorinlistoftodos]{todonotes}

\usepackage{hyperref}
\usepackage[normalem]{ulem}

\biboptions{sort,compress} 

\newcommand*\patchAmsMathEnvironmentForLineno[1]{%
  \expandafter\let\csname old#1\expandafter\endcsname\csname #1\endcsname
  \expandafter\let\csname oldend#1\expandafter\endcsname\csname end#1\endcsname
  \renewenvironment{#1}%
     {\linenomath\csname old#1\endcsname}%
     {\csname oldend#1\endcsname\endlinenomath}}%
\newcommand*\patchBothAmsMathEnvironmentsForLineno[1]{%
  \patchAmsMathEnvironmentForLineno{#1}%
  \patchAmsMathEnvironmentForLineno{#1*}}%
\AtBeginDocument{%
\patchBothAmsMathEnvironmentsForLineno{equation}%
\patchBothAmsMathEnvironmentsForLineno{align}%
\patchBothAmsMathEnvironmentsForLineno{flalign}%
\patchBothAmsMathEnvironmentsForLineno{alignat}%
\patchBothAmsMathEnvironmentsForLineno{gather}%
\patchBothAmsMathEnvironmentsForLineno{multline}%
}

\newcommand{\bstau}{\boldsymbol{\tau}}

\newcommand{\bstaurans}{\tilde{\boldsymbol{\tau}}^{rans}}

\newcommand{\bs}[1]{\boldsymbol{#1}}

\graphicspath{ {./Figures/} {./Figures/jinlong/} {./Figures/jianxun/} {./Figures/modes-duct} {./Figures/modes-pehill}}

\journal{Elsevier} 

\begin{document}

\begin{frontmatter}



  \title{Quantifying and Reducing Model-Form Uncertainties in Reynolds-Averaged Navier--Stokes
    Simulations: A Data-Driven, Physics-Based Bayesian Approach}



\author{H. Xiao\corref{corxh}}
\cortext[corxh]{Corresponding author. Tel: +1 540 231 0926}
\ead{hengxiao@vt.edu}

\author{J.-L. Wu\corref{corjl}}
\author{J.-X. Wang\corref{corjxw}}
\author{R. Sun\corref{corrs}}
\author{C. J. Roy\corref{corcjr}}

\address{Department of Aerospace and Ocean Engineering, Virginia Tech, Blacksburg, VA 24060, United States}

\begin{abstract}
  Despite their well-known limitations, Reynolds-Averaged Navier-Stokes (RANS) models are still the
  workhorse tools for turbulent flow simulations in today's engineering analysis, design and
  optimization. While the predictive capability of RANS models depends on many factors, for many
  practical flows the turbulence models are by far the largest source of uncertainty.  As RANS
  models are used in the design and safety evaluation of many mission-critical systems such as
  airplanes and nuclear power plants, quantifying their model-form uncertainties has significant
  implications in enabling risk-informed decision-making.  In this work we develop an open-box,
  physics-informed Bayesian framework for quantifying model-form uncertainties in RANS simulations.
  Uncertainties are introduced directly to the Reynolds stresses and are represented with compact
  parameterization accounting for empirical prior knowledge and physical constraints (e.g.,
  realizability, smoothness, and symmetry).  An iterative ensemble Kalman method is used to
  assimilate the prior knowledge and observation data in a Bayesian framework, and to propagate them
  to posterior distributions of velocities and other Quantities of Interest (QoIs). We use two
  representative cases, the flow over periodic hills and the flow in a square duct, to evaluate the
  performance of the proposed framework. Both cases are challenging for standard RANS turbulence
  models.  Simulation results suggest that, even with very sparse observations, the obtained
  posterior mean velocities and other QoIs have significantly better agreement with the benchmark
  data compared to the baseline results. At most locations the posterior distribution adequately
  captures the true model error within the developed model form uncertainty bounds.  The framework
  is a major improvement over existing black-box, physics-neutral methods for model-form uncertainty
  quantification, where prior knowledge and details of the models are not exploited.  This approach
  has potential implications in many fields in which the governing equations are well understood but
  the model uncertainty comes from unresolved physical processes. 
\end{abstract}


\end{frontmatter}





\section{Introduction}
\label{sec:intro}

\subsection{Model-form uncertainties in RANS-based turbulence modeling}

In Computational Fluid Dynamics (CFD), the Reynolds-Averaged Navier-Stokes (RANS) solvers are still
the workhorse tool for turbulent flow simulations in today's engineering analysis, design and
optimization, despite their well-known limitations, e.g., poor performance in flows with separation,
mean pressure gradient, and mean flow curvature~\cite{wilcox1998turbulence}. This is due to the fact
that high-fidelity models such as Large Eddy Simulation (LES) and Direct Numerical Simulation (DNS)
are still prohibitively expensive for engineering systems of practical interests. Moreover, in
engineering design and optimization, many simulations must be performed with short turn-around
times, which precludes the use of these high fidelity models.

The RANS equations employ a time- or ensemble-averaging process to eliminate temporal dependency for
stationary turbulence. The averaging leads to an unclosed correlation tensor, the Reynolds stress,
which needs to be modeled~\cite{wilcox1998turbulence,popebook}.  Turbulence modeling is a primary
source of uncertainty in the CFD simulations of turbulent flows.  Hundreds of RANS turbulence models
have been proposed so far.  Each has better performance in certain cases yet none is convincingly
superior to others in general.  This is due to the fact that the empirical closure models cannot
accurately model the regime-dependent, physics-rich phenomena of turbulent flows.  Predictions
obtained with any of these models have uncertainties that are difficult to quantify.  The model-form
uncertainties in RANS simulations originating from the turbulence models are the main focus of this
work.

\subsection{Model-Form uncertainty quantification: existing approaches}

A traditional approach for estimating RANS modeling uncertainties involves repeating the simulations
by perturbing the coefficients used in the turbulence models, or by using several different
turbulence models~\cite{roy2011comprehensive} (e.g., $k$--$\varepsilon$, $k$--$\omega$, and
eddy-viscosity transport models~\cite{wilcox1998turbulence}) and observe the sensitivity of the
Quantities of Interests (QoIs).  However, different models are often based on similar
approximations, and they are likely to share similar biases~\cite{saltelli15climate}. Consequently,
this \emph{ad hoc} model ensemble approach tends to underestimate of the uncertainty in the model.
In turbulence modeling, the Boussinesq assumption states that the Reynolds stress tensor is aligned
with and proportional to the local traceless mean strain rate tensor.  This assumption is shared by
all linear eddy-viscosity models that are commonly used in engineering practice, including the
$k$--$\varepsilon$, $k$--$\omega$, and eddy-viscosity transport models.

In their seminal work, Kennedy and O'Hagan~\cite{kennedy2001bayesian} developed a Bayesian
calibration approach that includes a model discrepancy term to account for Model-Form Uncertainty
(MFU). In this approach the MFU is quantified by parameterizing the difference between the
\emph{outputs} of the computational model and experimental observations as a stationary Gaussian
process whose hyperparameters can be inferred from data~\cite{kennedy2001bayesian}.  This framework
has been used in many applications, and a number of sophisticated variants have been developed,
e.g., by introducing non-stationary Gaussian processes to model the discrepancy~\cite{xiong2007non},
using multiplicative discrepancy term~\cite{huang2006sequential}, or using high-fidelity models and
field measurements to provide observation
data~\cite{conti2009gaussian,higdon2004co,huang2006sequential}.  While this
approach has had some success, the physics-neutral approach treats the entire numerical model as a
black box and does not exploit the prior information that often exists about the nature of the MFU
in a given model. Moreover, this framework addresses MFU only in terms of the QoIs, whereas the
modeling errors in RANS simulation arise specifically from the modeled Reynolds stress term.  Recent
work of Brynjarsdottir and O'Hagan~\cite{brynjarsdottir2014learning} emphasized the importance of
incorporating prior information, but they also highlighted the difficulties of enforcing prior
information in this black-box framework. Even a simple constraint such as zero-gradient boundary
condition on the discrepancy is challenging to enforce as shown
in~\cite{brynjarsdottir2014learning}. Realistic prior knowledge in engineering practice is generally
even more complicated.

Recently, several prominent groups in the CFD community (e.g., Moser and
co-workers~\cite{oliver2009uncertainty,oliver2011bayesian,cheung2011bayesian}, Iaccarino and
co-workers~\cite{gorle2013framework,gorle2014deviation,emory2013modeling,emory2011modeling,emory14estimate},
and Dow and Wang~\cite{dow11quanti}) have recognized the limitations of the black-box approach and
attempted to open the box by injecting the uncertainties locally into the closure models (i.e., not
on the model output directly). Research from these groups is reviewed in detailed below.  These
approaches have some similarities to earlier work of Berliner et
al.~\cite{wikle2001spatiotemporal,berliner2003physical} in the context of geophysical fluid
dynamics, where uncertainties were introduced to the discretized coefficients of the governing
geostrophic equations.

Moser and co-workers~\cite{oliver2009uncertainty,oliver2011bayesian,cheung2011bayesian} are the
first to explicitly point out and utilize the ``composite nature'' of the RANS equations. That is,
the equations are based on reliable theories describing conservation of mass, momentum, and energy,
but contain approximate embedded models to account for the unresolved or unknown physics, i.e., the
Reynolds stress terms. Based on this insight, they introduced a Reynolds stress discrepancy tensor
$\boldsymbol{\epsilon}$, which is added to the modeled Reynolds stress ($\bstaurans$) in the RANS
equations to account for the uncertainty due to the modeling of $\bstaurans$.  Stochastic differential
equations forced by Wiener processes are formulated for the discrepancy
$\boldsymbol{\epsilon}$. These equations are structurally similar to but simpler than the Reynolds
stress transport equations commonly used in turbulence
modeling~\cite[e.g.,][]{rodi75progress,wilcox1998turbulence}. Applications to plane channel flows
(where only the plane shear component of the Reynolds stress tensor is important) at various
Reynolds numbers have shown promising results, while extensions to general three-dimensional flows
are underway (Moser and Oliver, personal communication).

Iaccarino and
co-workers~\cite{gorle2013framework,gorle2014deviation,emory2013modeling,emory2011modeling,emory14estimate}
proposed a framework to estimate the model-form uncertainty in RANS modeling by perturbing the
Reynolds stress projections towards their limiting states within the physically realizable
range. Empirical indicator functions are used to ensure the spatial smoothness (i.e., spatial
correlation) of the perturbations in the physical domain, and to inject uncertainties only to the
regions where the baseline turbulence model is believed to perform poorly.  The novelty of their
framework is that both physical realizability and spatial correlations are accounted for, which are
two pieces of critical prior information in turbulence modeling.  Another advantage of their
framework is the moderate computational overhead, since only a few limiting states of the Reynolds
stresses are computed.  On the other hand, it should be noted that the obtained scattering of the
states can only serve as an empirical estimation of the uncertainties, and are not guaranteed to
cover the truth. While the true Reynolds stress is a convex linear combination of the Reynolds
stresses in the limiting states, the true velocities or other QoIs are not necessarily linear
combinations of their respective limiting states.

Dow and Wang~\cite{dow11quanti} quantified model-form uncertainties in the $k$--$\omega$ model by
finding the eddy viscosity field that minimizes the misfit in the computed velocity field compared
to the DNS data. While their approach has some similarities with that of Iaccarino et al., the most
notable difference is that uncertainties are injected to the eddy viscosity and not to the Reynolds
stresses directly.  Another key difference is that they used DNS data, while Iaccarino et al. did
not and instead focused only on forward propagation of uncertainties in the Reynolds stress.

Duraisamy et al.~\cite{tracey2015machine, parish2016paradigm, singh16using}, on the
  other hand, introduced uncertainties as full-field multiplicative discrepancy term $\beta$ in the
  production term of the transport equations of turbulent quantities (e.g., $\tilde{\nu}_t$ in the
  SA model and $\omega$ in the $k$--$\omega$ models).  Full-field DNS data or sparse data from
  experimental measurements were used to calibrate and infer uncertainties in this term. It is
  expected that the inferred discrepancy field can provide valuable insights to the development of
  turbulence models. They also suggested the possibility of extrapolating the learned discrepancy
  fields to similar flows via machine learning techniques.

In summary, the CFD community has recognized the advantages of open-box approaches for quantifying
model-form uncertainties in RANS simulations, and promising results have been obtained.  However, much
work is still needed.

\subsection{Objective and Novelty of the Present Work}
\label{sec:objective}
In this work, we focus on a scenario where a limited amount of data (usually from measurements at a
few locations) is available. This is often the case when CFD is used in practical applications in
conjunction with experimental data to provide predictions.  Examples include prediction of flows in
a wind farm and atmospheric pollutant dispersion in a city~\cite{gorle2013framework}. Built on
existing insights and experiences in the literature, the objective of this work is to develop a
rigorous, open-box, physics-informed framework for quantifying model-form uncertainties in RANS
simulations. Compared to the pioneering framework of Iaccarino et
  al.~\cite{gorle2013framework,gorle2014deviation,emory2013modeling,emory2011modeling,emory14estimate}
  where the model-form uncertainty in RANS simulations was estimated by perturbing the Reynolds
  stresses towards their three limiting states, the novelty of our approach is that an
ensemble-based Bayesian inference method is used to incorporate all sources of available
information, including empirical prior knowledge, physical constraints (e.g., realizability,
smoothness, and symmetric), and available observation data.

This work aims to quantify and reduce the model form uncertainty by utilizing both
  state-of-the-art statistical inference techniques and domain knowledge in turbulence modeling. As
  a first step, we focus on an idealized scenario where model-form uncertainty is the dominant
  source of uncertainty, and the coupling with other uncertainties, e.g., model input uncertainty
  and numerical uncertainty, is not considered.

The proposed framework has been evaluated on two canonical flows, the flow over periodic
  hills and the flow in a square duct, in the present work. Further application to a more
  complicated, three dimensional flow of critical relevance to aerospace engineering, i.e., the flow
  over a wing--body junction, has also been explored and presented in a separate
  work~\cite{wu2016wing}. While the authors believe that the present contribution is novel and
  represents an advancement over the state of the art, we expect significant challenges that need to
  be addressed before the proposed approach can be extended to industrial flows, e.g., the flows past
  an aircraft or in a gas turbine.

The rest of the paper is organized as follows. The model-form uncertainty quantification framework
is introduced in Section~2, and numerical implementation details are given in Section~3. Numerical
results for two application cases, the flow over periodic hills and the flow in a square duct, are
presented in Section~4 to assess the merits and limitations of the developed framework. The
success, limitations, practical significance, and possible extensions of the proposed method are
further discussed in Section~5. Finally, Section~6 concludes the paper.

\section{Proposed Framework}
\label{sec:method}

\subsection{Prior Knowledge in RANS Modeling} 
\label{sec:prior}

\begin{figure}[htbp]
  \centering
  \includegraphics[width=0.75\textwidth]{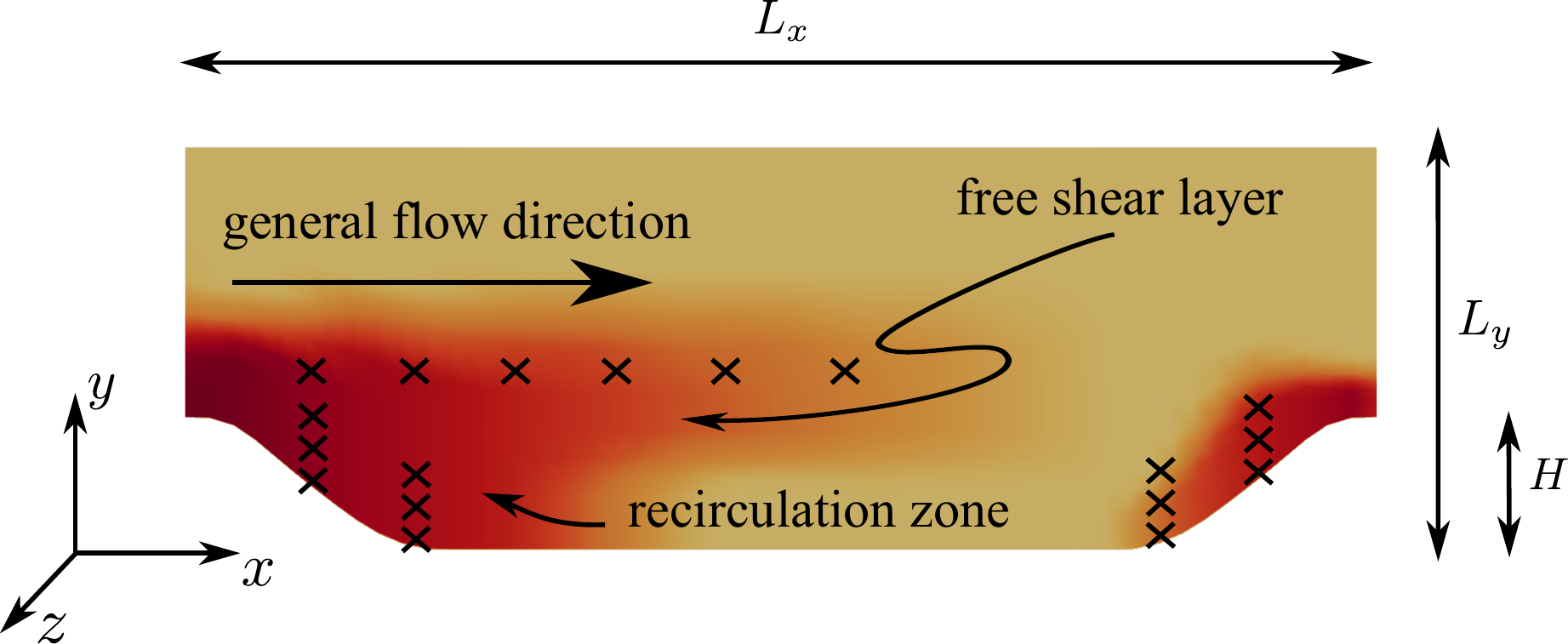}
  \caption{Domain shape for the flow over periodic hills. The $x$-, $y$- and $z$-coordinates are
    aligned with streamwise, wall-normal and spanwise directions, respectively. All dimensions are
    normalized with $H$ with $L_x/H=9$, $L_y/H=3.036$. The contour shows the variance field
    $\sigma(x)$, where darker color represents the larger variance.  The locations where velocities
    are observed are indicated as crosses ($\times$).}
  \label{fig:domain_pehill}
\end{figure}

An important feature of the proposed framework is the explicit, straightforward representation of
prior knowledge in a Bayesian inference framework. As such, we summarize the prior knowledge in
RANS-based turbulent flow simulations below, some of which has been reviewed in
Section~\ref{sec:intro}:
\begin{enumerate}[(a)]
\item \emph{Composite model:} The uncertainties in the modeled Reynolds stresses are the main source
  uncertainties in the RANS model predictions~\cite{oliver2009uncertainty}.
\item \emph{Physical realizability:} The true Reynolds stress at any point in the domain resides in
  a subspace of a six-dimensional space~\cite{tennekes1972first,emory2011modeling}.
\item \emph{Spatial smoothness:} The Reynolds stress field usually has smooth spatial distributions
  except across certain discontinuous features (e.g., shocks and abrupt changes of geometry).
\item \emph{Problem-specific prior knowledge:} There are some well-known scenarios where eddy
  viscosity models are expected to perform poorly as enumerated above, e.g., flow separation, mean
  flow curvature. Taking the flow over periodic hills as shown in Fig.~\ref{fig:domain_pehill} for
  example, the flood contour indicates typical prior knowledge of the relative magnitude of the
  Reynolds stress discrepancies in each region, i.e., the regions with recirculation, non-parallel
  free-shear flow, and the strong mean flow curvature have larger discrepancies.
\end{enumerate}

\subsection{Representations of Prior Knowledge in the Modeling Framework}
\label{sec:pripr-rep}

In light of the prior knowledge presented above and based on the existing methods in the
literature~\cite{oliver2009uncertainty,emory2011modeling,dow11quanti}, we make the following
modeling choices to represent the prior knowledge.

\subsubsection{Composite model}

The true Reynolds stress $\bstau$ is modeled as a random field of symmetric tensors with $\bstaurans$
as its deterministic mean field, where $\bstaurans$ is the Reynolds stress field given in the
baseline RANS simulation whose model-form uncertainty is to be quantified.\footnote{We use
  $~\tilde{}~$ to emphasize the fact that $\bstaurans$ is a deterministic field, which is in contrast
  to the random field $\bstau$.}

\subsubsection{ Physical realizability of Reynolds stresses}
\label{sec:phys-rep}

\begin{figure}[!htb]
  \centering
   \includegraphics[width=0.9\textwidth]{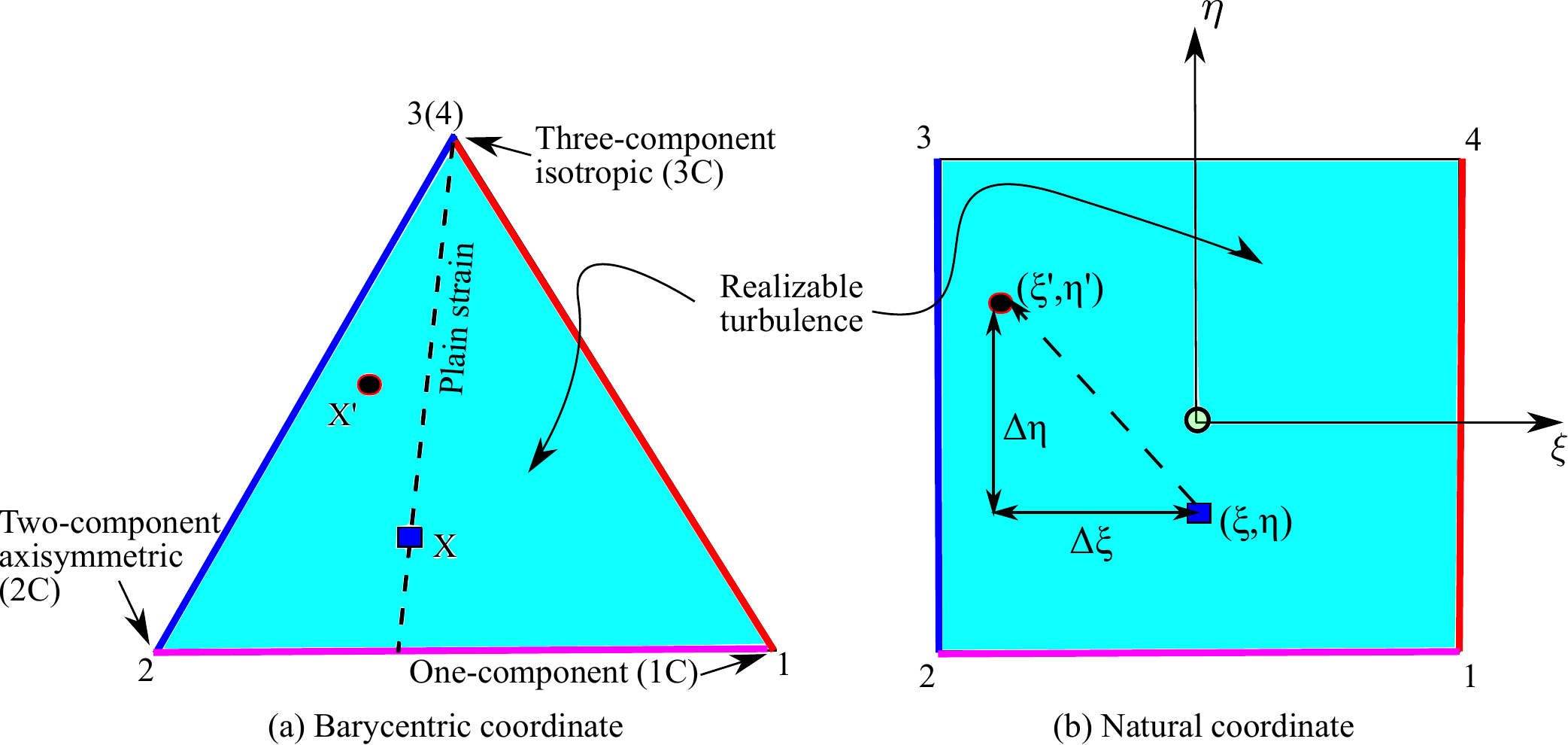}
   \caption{Mapping between Barycentric coordinates and natural coordinates, transforming the
     Barycentric triangle that encloses all physically realizable
     states~\cite{banerjee2007presentation,emory2013modeling} to a square via standard finite
     element shape functions (detailed in \ref{app:mapping}). Corresponding edges in the two
     coordinates are indicated with matching colors. The singular point 3(4) in the Barycentric
     coordinate, which maps to the edge 3--4 in the natural coordinate, does not pose any practical
     difficulties.}
  \label{fig:bary}
\end{figure}

To ensure physical realizability of its realizations, the value of the Reynolds stress field
$\bstau$ at any given location $x$ is projected onto a space with six physically meaningful
dimensions via the following eigen-decomposition~\cite{emory2011modeling,gorle2013framework}:
\begin{equation}
  \label{eq:tau-decomp}
  \boldsymbol{\tau} = 2 k \left( \frac{1}{3} \mathbf{I} +  \mathbf{a} \right)
  = 2 k \left( \frac{1}{3} \mathbf{I} + \mathbf{V} \Lambda \mathbf{V}^T \right)
\end{equation}
where $k$ is the turbulent kinetic energy, $\mathbf{I}$ is the second order unit tensor,
$\mathbf{a}$ is the anisotropy tensor, $\mathbf{V} = [\mathbf{v}_1, \mathbf{v}_2, \mathbf{v}_3]$, and $\Lambda =
\textrm{diag}[\lambda_1, \lambda_2, \lambda_3]$ are its orthonormal eigenvectors and eigenvalues,
respectively, with $\lambda_1+\lambda_2+\lambda_3=0$.  This decomposition transforms the Reynolds
stress to a space represented by six variables with clear physical interpretations: magnitude
(represented by the turbulent kinetic energy $k$, which must be non-negative), shape (represented by
two scalars $\lambda_1$, $\lambda_2$), and orientation (represented by three mutually orthonormal
vectors\footnote{They can be considered as the three orthogonal axes of an ellipsoid, and thus the
  three vectors have three degrees of freedom in total, i.e., its orientation in three-dimensional
  space.} $\mathbf{v}_1$, $\mathbf{v}_2$, and $\mathbf{v}_3$) of the Reynolds stress
tensor~\cite{banerjee2007presentation,gorle2013framework}.  Further, $\lambda_1$, $\lambda_2$, and
$\lambda_3$ are transformed to the Barycentric coordinates ($C_1$, $C_2$, $C_3$), with $C_1 + C_2 +
C_3 = 1$, and subsequently to the natural coordinates ($\xi$, $\eta$). With the mapping from
Barycentric coordinates to natural coordinates (see Fig.~\ref{fig:bary}), the physically realizable
turbulent stresses enclosed in the Barycentric triangle (panel a) are transformed to a square (panel
b), i.e., $\{ (\xi, \eta) \, | \, \xi \in [-1, 1] , \eta \in [-1, 1] \}$, which is more convenient
for parameterization. Details of the mapping are presented in~\ref{app:mapping}. In summary, we
transform the Reynolds stress tensor to six physical dimensions denoted as $(k, \xi, \eta,
\mathbf{v}_1, \mathbf{v}_2, \mathbf{v}_3)$.  All mappings involved are linear and invertible except
for a trivial singular point in $(C_1, C_2, C_3) \mapsto (\xi, \eta)$.

After the mapping of Reynolds stress $\bstaurans$ to the physically meaningful dimensions, i.e.,
$k$, $\xi$, $\eta$, uncertainties are injected to the projected space on these variables. This is
achieved by modeling the corresponding truths $k(x)$, $\xi(x)$, and $\eta(x)$ as random fields with
$\tilde{k}^{rans}(x)$, $\tilde{\xi}^{rans}(x)$, and $\tilde{\eta}^{rans}(x)$ as
priors. Specifically,
\begin{subequations}
    \label{eq:delta-def}
  \begin{align}
  \log  k(x) & =  \log \tilde{k}^{rans}(x) + \delta^k(x)  \label{eq:kdelta} \\
    \xi (x) & = \tilde{\xi}^{rans}(x) + \delta^\xi(x)  \\
    \eta(x) & = \tilde{\eta}^{rans}(x) + \delta^\eta(x)
  \end{align}
\end{subequations}
where the spatial coordinate $x$ is the index of the random fields.  Note that the logarithmic
discrepancy of the turbulent kinetic energy $k$ is modeled in Eq.~(\ref{eq:kdelta}) to ensure the
non-negativity of $k$.

The realizability in this framework is ensured by bounding the perturbed anisotropy ($\eta$,
  $\xi$) within the square $[-1, 1] \times [-1, 1]$ in the $\xi$--$\eta$ plane as shown in
  Fig. 2b. Any perturbed state outside this range will be bounded to the edge of the square, which
  is admittedly an ad hoc modeling choice.  As a result, the prior may become non-Gaussian and the
  perturbation sample may deviate from zero-mean if a large number of perturbations are bounded.
  However, note that for a Gaussian prior the percentage of out-of-bound points can be estimated,
  and thus the variance of the perturbation can be controlled straightforwardly given an allowable
  ratio of out-of-bound points. This is one of the advantages of mapping the Barycentric triangle to
  the square before introducing perturbation as opposed to directly perturbing the baseline within
  the Barycentric triangle.

The bounding scheme for ensuring realizability can distort the distribution of the sampled
  Reynolds stresses from the specified prior.  Specifically, when the baseline Reynolds stresses are
  located near the realizability boundaries, e.g., the top vertex of the Barycentric triangle for
  points near walls, the bounding can cause the sample distribution to become truncated Gaussian.
  However, note that the tail truncation and the associated probability mass concentration near the
  boundaries are caused by the bounding procedure to ensure realizability, regardless of whether
  Barycentric coordinates or natural coordinates are used. This issue is further investigated in two
  follow-on studies~\cite{RMT1, RMT2}, where we proposed a random matrix approach which directly
  samples a maximum entropy distribution defined on the set of positive semidefinite matrices, and
  the artificial probability mass concentration described above is avoided.  However, note that the
  approximate Bayesian inference method (i.e., the ensemble Kalman method) used in this work is not
  sensitive to the prior. From a practical point of view, the prior can be alternatively interpreted
  as an ``initial guess'' used in optimization~\cite[e.g.,][]{parish2016paradigm}.

Perturbing the orientations of the modeled Reynolds stress tensor can potentially cause instability
in the RANS momentum equation. Consistent with the work of Iaccarino et al., we focus on the
magnitude ($k$) and the shape ($\lambda_1$ and $\lambda_2$, or equivalently the natural coordinates
$\xi$ and $\eta$) of the Reynolds stress tensor $\bstau$, and do not introduce uncertainties into
the orientations~($\mathbf{v}_1, \mathbf{v}_2, \mathbf{v}_3$). Consequently, the assumed uncertainty
space of Reynolds stresses may not contain the truth because the true Reynolds stresses are likely
to have different orientations from those of the RANS predictions. The implications of this fact
will be further discussed in Section~\ref{sec:pehill-res}.

\subsubsection{Spatial smoothness of Reynolds stress distribution}
\label{sec:smooth-rep}

To ensure spatial smoothness and to reduce the dimension of the uncertainty space, the random fields
to be inferred, i.e., $\delta^k$, $\delta^\xi$, and $\delta^\eta$, are projected to a deterministic
functional basis set \{$\phi_i(x)$\}. That is,
  \begin{subequations}
    \label{eq:delta-proj}
  \begin{align}
    \delta^k(x, \theta^k) = \sum_{i=1}^\infty \omega^k_{i} |_{\theta^{k}} \; \phi_i (x) \\
    \delta^\xi(x, \theta^\xi) = \sum_{i=1}^\infty \omega^{\xi}_{i} |_{\theta^{\xi}} \; \phi_i (x) \\
    \delta^\eta(x, \theta^\eta) = \sum_{i=1}^\infty \omega^\eta_{i} |_{\theta^{\eta}} \; \phi_i (x)
  \end{align}
\end{subequations}
where the coefficients of the $i^{th}$ mode $\omega^k_{i}$, $\omega^\xi_{i }$, and $\omega^\eta_{i}$
are random variables\footnote{Throughout the manuscript, subscripts denote indices, and
  superscripts indicate explanation of the variable. For example, $\omega^k_{i}$ is the coefficient
  for the $i^{th}$ mode in the expansion of the discrepancy field $\delta^k$ for the turbulent
  kinetic energy $k$. Tensors are denotes in bold (e.g., $\bstau$) and not with index notation.}
depending on the realized outcome of $\theta^{k}$, $\theta^{\xi}$, and $\theta^{\eta}$,
respectively, and $\phi_i(x)$ are deterministic spatial basis functions.  An orthogonal basis set is
chosen in this work as will be detailed below, but the orthogonality is not mandatory.

\emph{Remarks:} The mapping in Section~\ref{sec:phys-rep} involves linear transformation of the
Reynolds stress at a given point to physical variables, which ensures the physical realizability of
the Reynolds stresses in the prior. The orthogonal projection in Section~\ref{sec:smooth-rep} aims
to represent the spatial distribution function on a basis set in a compact manner, which ensures
spatial smoothness and reduces the uncertainty dimensions of $\bstau(x)$.

\subsubsection{Representation of problem-specific prior knowledge}
\label{sec:pripr-prob-spec}

Finally, problem-specific knowledge is encoded in the choice of basis set \{$\phi_i$\}. Here we will
use the flow over periodic hills as example to illustrate the representation of the problem-specific
prior knowledge.

We model the prior of the discrepancies $\delta^k$, $\delta^\xi$ and $\delta^\eta$ as zero-mean
Gaussian random fields (also known as Gaussian processes) $\mathcal{GP}(0, K)$, where
  \begin{equation}
  \label{eq:gp-kernel}
  K(x, x') = \sigma(x)  \sigma(x') 
  \exp \left( - \frac{|x - x'|^2}{l^2}  \right)
\end{equation}
is the kernel indicating the covariance at two locations $x$ and $x'$. The variance $\sigma(x)$ is a
spatially varying field specified (see the flood contour in Fig.~\ref{fig:domain_pehill}) to
reflect the prior knowledge that large discrepancies in modeled Reynolds stress are expected in
certain regions.  The correlation length scale $l$ can be specified based on the local turbulence
length scale, but is taken as constant in this work for simplicity.

The orthogonal basis functions $\phi_i(x)$ in Eq.~(\ref{eq:delta-proj}) take the form ${\phi}_i(x) =
\sqrt{\hat{\lambda}_i} \hat{\phi}_i(x) $, where $\hat{\lambda}_i$ and $\hat{\phi}_i(x)$ are
eigenvalues and eigenfunctions, respectively, of the kernel $K$ in Eq.~(\ref{eq:gp-kernel}) computed
from the Fredholm integral equation~\cite{le2010spectral}:
\begin{equation}
\label{eq:kl}
\int K(x, x') \hat{\phi}(x') \, dx'  = \hat{\lambda} \hat{\phi}(x) \, .
\end{equation}
With this choice of basis set the expansions in Eq.~(\ref{eq:delta-proj}) for the fields $\delta^k$,
$\delta^\xi$ and $\delta^\eta$ become Karhunen--Loeve (KL) expansions~\cite{le2010spectral}, such
that $\omega^k_{i}$, $\omega^\xi_{i}$, and $\omega^\eta_{i}$ are uncorrelated random variables with
zero means and unit variances.

\paragraph{Remarks} The Gaussian process and KL expansions are intentionally presented in this
Section to emphasize the fact that they are our specific choices for this problem and prior
knowledge only. The optimal choice of basis set depends on the specific characteristics (e.g.,
smoothness, compactness of support) of the prior.  Other functional basis sets, including
wavelets~\cite{daubechies1988orthonormal} or radial basis functions~\cite{buhmann2003radial}, will
be explored in future work.

\subsection{Inverse modeling based on an iterative ensemble Kalman method}
\label{sec:enkf}

After the transformations above, the Reynolds stress random field $\bstau(x)$ is parameterized by
the coefficients $\omega^k_{i}$, $\omega^\xi_{i}$, and $\omega^\eta_{i}$ in
Eq.~(\ref{eq:delta-proj}), which are truncated to $m$ modes and written in a stacked vector form as
follows\footnote{It is trivial for each variable to have a different number of modes, but this
  possibility is omitted here to simplify notation.}:
\begin{equation}
  \label{eq:omega}
  \bs{\omega} \equiv [\omega^k_{1}, \, \omega^\xi_{1}, \,  \omega^\eta_{1}, \;
 \omega^k_{2},  \, \omega^\xi_{2}, \, \omega^\eta_{2}, \; \cdots,  \; 
  \omega^k_{m}, \, \omega^\xi_{m}, \, \omega^\eta_{m}]
\end{equation}

\begin{figure}[!htb]
\centering
\includegraphics[width=0.95\textwidth]{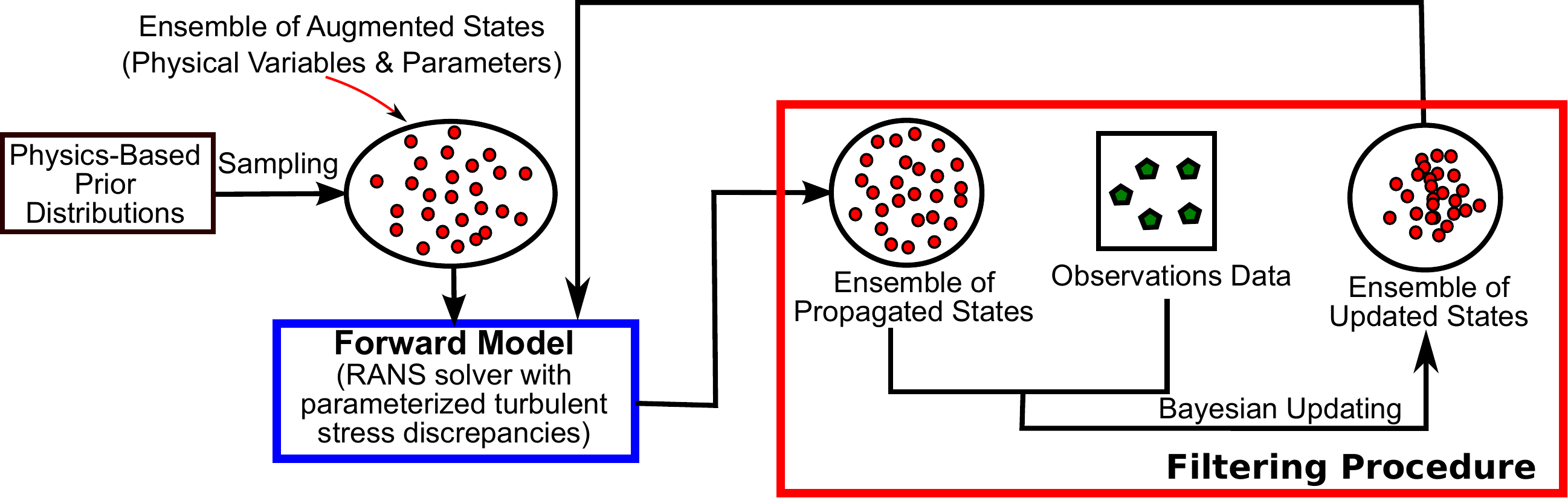}
\caption{Inference of coefficients in the parameterized model discrepancies (e.g., discrepancies in
  RANS modeled Reynolds stresses) using an iterative ensemble Kalman inversion method. This approach combines
  prior knowledge of a given problem and available data to quantify and reduce model-form
  uncertainty.
  \label{fig:enkf}}
\end{figure}

We employ an \emph{iterative, ensemble-based} Bayesian inference method~\cite{iglesias2013ensemble}
to combine the prior knowledge as represented above and the available data to infer the distribution
of $\bs{\omega}$. This method is closely related to ensemble filtering methods (e.g., ensemble
Kalman filtering), which are a class of standard data assimilation techniques commonly used in
numerical weather forecasting~\cite{evensen2009data}.  An overview of the ensemble Kalman method
based inverse modeling procedure is presented in Fig.~\ref{fig:enkf}.  In the iterative ensemble
method, the state of the system $\mathbf{x}$ is defined to include both the physical variables
(i.e., velocity field $\mathbf{u}$) and the unknown coefficients $\bs{\omega}$, i.e., $\mathbf{x}
\equiv [ \mathbf{u}, \bs{\omega}]^T$. This is called ``state
augmentation''~\cite{iglesias2013ensemble}.  One starts with an ensemble of states
$\{\mathbf{x}_j\}_{j=1}^N$ drawn from their prior distributions.  During each iteration, all samples
in the ensemble are updated to incorporate the observations through the following procedure:
\begin{enumerate}[(a)]
\item reconstruction of Reynolds stresses from the coefficients $\bs{\omega}$,
\item computation of velocity fields from the given Reynolds stress fields by solving the RANS
  equations (implemented as forward model tauFoam, detailed in Section~\ref{sec:implement}), and
\item a Kalman filtering procedure to assimilate the velocity observation data to the computed
  states, leading to an updated ensemble.
\end{enumerate}
The updating procedure is repeated until the ensemble is statistically converged. The convergence is
achieved when the two-norm of the misfit between the predictions and the observations falls below
the noises level of the observations~\cite{iglesias2013ensemble}.  The converged ensemble is
considered a sample-based representation of the posterior distribution of the system state, from
which the mean, variance, and higher moments can be computed. The algorithm of the inversion scheme
is presented in~\ref{app:enkf}, and further details can be found in~\cite{iglesias2013ensemble}.

The noises added to the observations represent a combination of measurement errors and process
errors~\cite{dennis2006estimating}. The former is likely to be negligible for DNS data. However, the
latter can be significant and is used to account for the fact that the observed system and the
system described in the numerical model can have different dynamics. From a Bayesian perspective,
adding the process noise allows the likelihood and the prior distribution to have overlap in their
supports and thus be able to reconcile with each other in the inference procedure.  As long as the
chosen noise level $\sigma_{obs}$ is larger than a threshold (1\% of truth in this work), the
inferred posterior means are not sensitive to this parameter. See further discussion
in~\cite[e.g.,][]{iglesias2013ensemble}.

An important property of the iterative ensemble Kalman method is that the posterior ensembles and
its mean all lie in the linear space $\mathcal{A}$ spanned by the prior ensemble
$\{\mathbf{x}_j\}_{j=1}^N$. In essence, this scheme attempts to search the space $\mathcal{A}$ to
find the optimal solution that minimizes the misfit between the posterior mean and the observations,
accounting for the uncertainties in both~\cite{iglesias2013ensemble}. As with many inverse problems,
this problem is intrinsically ill-posed. Specifically, because of the sparseness of the observation
(the scenario of concern in our work), the amount of data is usually not sufficient to constrain the
uncertainties in the states, which include the model discrepancies as components.  The forward model
essentially provides the regularization of the ill-posedness with its physical representation of the
system dynamics.

The ensemble Kalman-based uncertainty quantification scheme used here is an approximate Bayesian
method, and is computationally cheaper than the exact Bayesian scheme based on Markov Chain Monte
Carlo sampling. It is not expected to give posterior distributions with comparable accuracy to those
obtained from exact Bayesian schemes~\cite{law12evaluate}. This limitation will be further discussed
in Section~\ref{sec:discuss}.

\subsection{Summary of the algorithm in the proposed framework}
\label{sec:algo}

In summary, the overall algorithm of the proposed framework for quantifying and reducing
uncertainties in a RANS simulation is presented as follows.

\begin{enumerate}
\item Perform the baseline RANS simulation to obtain the velocity $\tilde{\bs{u}}^{rans}(x)$ and
  Reynolds stress $\bstaurans(x)$.
\item Perform the transformation $\bstaurans \mapsto (\tilde{k}^{rans}, \, \tilde{\xi}^{rans}, \,
  \tilde{\eta}^{rans})$.  
\item Compute KL expansion to obtain basis set $\{\phi_i(x)\}_{i=1}^m$, where $m$ is the number of
  modes retained.
\item Generate initial prior ensemble of coefficient vectors $\{ \bs{\omega}_j \}_{j=1}^N$, where
  $N$ is the ensemble  size.
\item Use iterative scheme shown in Fig.~\ref{fig:enkf} to obtain the posterior ensemble of the state
  distribution. Specifically, in each iteration do the following:
     \begin{enumerate}
     \item Recover the discrepancy fields $\delta^k$, $\delta^\xi$, and $\delta^\eta$ from the
       coefficient $\{ \bs{\omega}_j \}_{j=1}^N$ in the current state and the basis functions via
       Eq.~(\ref{eq:delta-proj}), and obtain realizations of $k$, $\xi$, and $\eta$ from
       Eq.~(\ref{eq:delta-def}) for each sample in the ensemble.
     \item Obtain Reynolds stress ensembles $\{\bstau_j \}_{j=1}^N$ via mapping $(k, \xi, \eta)
       \mapsto \bstau$.
     \item For each sample in the ensemble $\{\bstau_j \}_{j=1}^N$, solve the RANS equations for
       velocity field $\bs{u}_j$ with given Reynolds stress field $\bstau_j$.
     \item Compare the ensemble mean with velocity observations, and use the Kalman filtering procedure
       to correct the augmented system state ensemble $\{\mathbf{x}_j \}_{j=1}^N$, where
       $\mathbf{x}_j = [\mathbf{u}_j, \bs{\omega}_j]^T$. The updated coefficient vector ensemble
       $\bs{\omega}_j$ is thus obtained as part of the system state ensemble.
     \item Stop if statistical convergence of the ensemble as defined in Section~\ref{sec:enkf} is
       achieved.
  \end{enumerate}
\end{enumerate}

\section{Implementation and Numerical Methods}

\label{sec:implement}

The uncertainty quantification framework including the mapping of Reynolds stresses and the
iterative ensemble Kalman method is implemented in Python, which interfaces with RANS models and the
KL expansion procedures to form the complete framework.  The package UQTk developed by Sandia
National Laboratories is used to perform the KL expansions~\cite{uqtk}.  Two types of RANS solvers
are used in this framework, a conventional baseline RANS solver \texttt{simpleFoam} and a forward
RANS solver \texttt{tauFoam} which computes velocity field with a given Reynolds stress field. Both
solvers are described as below.

The baseline simulation uses a built-in RANS solver \texttt{simpleFoam} in OpenFOAM for
incompressible, steady-state turbulent flow simulations. OpenFOAM (for ``Open source Field Operation
And Manipulation'') is an open-source, general-purpose CFD platform based on finite-volume
discretization. The platform consists of a wide range of solvers and post-processing utilities.  The
SIMPLE (Semi-Implicit Method for Pressure Linked Equations) algorithm~\cite{patankar72} is used to
solve the coupled momentum and pressure equations. Collocated grids are used, and the Rhie and Chow
interpolation is used to prevent the pressure--velocity decoupling~\cite{rhie83}.  Second-order
spatial discretization schemes are used to solve the equations on an unstructured, body-fitting
mesh.  Given the specification of the flow including initial conditions (to start the iteration),
boundary conditions, geometry, and the choice of turbulence model, the \texttt{simpleFoam} solver
computes the velocity field along with Reynolds stresses by solving the RANS equations as well as
the equations for the turbulence quantities (e.g., turbulent kinetic energy $k$ and the rate of
dissipation $\varepsilon$ for $k$--$\varepsilon$ models).  We choose the Launder--Sharma low
Reynolds number $k$--$\varepsilon$ model~\cite{launder-sharma74} in the baseline
simulations. Accordingly, the meshes are refined wall-normal direction near the wall to resolve the
boundary layer. This is to avoid the complexity of using wall-functions, which is in consistent with
the work of Emory et al.~\cite{emory2011modeling}.  As can be seen in the overall algorithm
presented in Section~\ref{sec:algo}, for each uncertainty quantification case the baseline
simulation is performed only once.

The forward RANS model \texttt{tauFoam} is invoked repeatedly in the Bayesian inference
procedure. This solver is adopted from and similar to \texttt{simpleFoam} except that it computes the
velocity directly with a \emph{given Reynolds stress field}.  There is no need to specify a
turbulence model and or to solve the equations for turbulence quantities, since the Reynolds stress
is given. Moreover, as the forward RANS simulations are initialized with the converged baseline
solutions, the number of iterations needed to achieve convergence is much smaller than that in the
baseline simulation. As a result, the computational cost for each call of the forward RANS model
\texttt{tauFoam} is much lower than that of the conventional RANS solver \texttt{simpleFoam}. In the
simulations presented below, the forward RANS simulations need only 10\%
of the computational cost as that of the baseline simulation to achieve the same residual.

\section{Numerical Simulations}
\label{sec:simulations}

Two canonical flows, the flow in a channel with periodic constrictions (periodic hills) and the
fully developed turbulent flow in a square duct, are chosen to evaluate the performance of the
proposed framework. The periodic hill flow features a recirculation zone formed by a forced
separation, a strong mean flow curvature due to the domain geometry, and a shear layer that is not
aligned with the overall flow direction. All these features are known to pose challenges for
turbulence modeling. The square duct flow is characterized by a secondary flow pattern in the plane
perpendicular to the main flow. The in-plane secondary flow is driven by the imbalance in the normal
components of the Reynolds stress tensor, which cannot be captured by models with isotropic eddy
viscosity turbulent models including most of the widely used models such as $k$--$\varepsilon$,
$k$--$\omega$, and eddy viscosity transport models. The two challenging cases are chosen to
demonstrate the capability of the proposed framework in quantifying and reducing uncertainties in
the RANS model predictions by incorporating sparse observations.

\subsection{Flow over Periodic Hills}

\subsubsection{Case setup}

The periodic hill flow is widely used in the CFD community to evaluate the performance of turbulence
models due to the availability of experimental and numerical benchmark data~\cite{breuer2009flow}.
The geometry of the computational domain and the coordinate system are shown in
Fig.~\ref{fig:domain_pehill}.  The Reynolds number based on the crest height $H$ and the bulk flow
velocity $U_b$ at the crest is $Re_b = 2800$. Periodic boundary conditions are applied in the
streamwise ($x$) direction, and non-slip boundary conditions are applied at the walls. The mean flow
is two-dimensional, and thus the spanwise ($z$) direction is not considered for the RANS simulations.

The mesh and computational parameters used in the uncertainty quantification procedure are presented
in Table~\ref{tab:paraDA}.  Despite the coarse meshes, the walls are adequately resolved in both
cases as required by the Lauder--Sharma turbulence model~\cite{launder-sharma74}.  The distance
between the center of the first cell and the wall is smaller than 1 in most regions for periodic hill
case and 0.7 for the square duct case. Parameters of the meshes for both cases are shown in
Table~\ref{tab:paraDA}.

The uncertainties in $\xi$, $\eta$, $k$ are all considered, and thus
$\delta^\xi$, $\delta^\eta$ and $\delta^k$ are all random fields. This choice is based on our prior
knowledge that both the Reynolds stress anisotropy (indicated by the shape $\bstau$, or
equivalently, $\xi$ and $\eta$) and turbulent kinetic energy $k$ predicted by the RANS model are
biased, and both are important for the accurate prediction of the flow behavior.  The length scale
parameter $l$ is chosen according to the approximate length scale of the flow, which can be obtained
either from our physical understanding of the flow or, if that is not available, from the baseline
RANS simulation.  Velocity observations are generated by adding Gaussian random noises with standard
deviation $\sigma_{obs}$ to the truth from DNS data. Specifically, the observations used in each
iteration are independent realizations from a Gaussian distribution whose mean is the truth and
standard deviation $\sigma_{obs}$ is 10\% of the true mean value.  The noises at different locations
are uncorrelated. The observation points are arranged so that they are closer in regions where the
spatial changes of the flow are more rapid (the recirculation zone leeward of the hill and the
reattached flow region windward of the hill), and are further apart in the free shear region
downstream of the hill crest. This arrangement of observations is expected in actual experiments.
The ensemble usually converges in approximately 10 iterations.  For all cases presented in
  this work, 60 samples are used in the ensemble. We have performed detailed sensitivity studies on
  the ensemble size, and it was found that the inferred velocities and QoIs do not vary if more than
  30 samples are used. This finding is consistent with earlier studies when EnKF was used in data
  assimilations in applications such as weather forecasting~\cite{houtekamer2009model}. The
computational cost of the proposed procedure is further discussed in Section~\ref{sec:cost}.

The non-stationary Gaussian process models for $\delta^\xi$, $\delta^\eta$ and $\delta^k$ share the
same variance field $\sigma(x)$, which are shown as flood contour in Fig.~\ref{fig:domain_pehill}.
Design of the variance field is strictly based on physical prior knowledge as described in
Section~\ref{sec:pripr-prob-spec}, and does not take the DNS data into account, since the complete
field of the true Reynolds stresses are rarely known in practical applications. Specifically, the
variance fields of the priors as shown in Figs.~\ref{fig:domain_pehill} and \ref{fig:domain_duct}
are constructed by superimposing a constant background value $\sigma_0$ and a spatially varying
field $\sigma_{local}(x)$, i.e., $\sigma(x) = \sigma_{0} + \sigma_{local}(x)$.  For the periodic
hill case, the background $\sigma_{0}$ is set to be 0.2.  To obtain the field $\sigma_{local}(x)$,
we specify that $\sigma_{local} = 0.5$ at the following locations, where RANS predictions are
considered less reliable: (1) the hill crest, (2) the center of the recirculation region, (3) the
windward side of the hill, and (4) the free-shear layer downstream the crest. Interpolations based
on radial basis functions with exponential kernels are used to obtain $\sigma_{local}(x)$ at other
locations. As such, its value decays to zero far away from the locations specified above. The length
scale of the basis functions is estimated based on the characteristic length of the mean flows,
which is chosen as the hill height $H$.

The first sixteen modes obtained from the KL expansion are used to reconstruct the discrepancy
field. The number of modes retained is chosen such that the reconstructed field has at least 80\% of
the total variance of the original random field.  A rule of thumb is that a coverage ratio of 80\%
is adequate for a faithful representation. Increasing the number of modes increases the difficulty
of the inference and may lead to deteriorated results for a given amount of observation data.

Parameter sensitivity analysis has been conducted to ensure that reasonable variations of the
computational parameters above do not lead to significantly different results or conclusions.  In
particular, we have shown in a follow-on study~\cite{mfu2} that even in the complete absence of
prior knowledge (i.e., a constant variance field $\sigma(x)$), the inferred velocities are still
significantly improved, albeit slightly less so than that with an informative prior.  Specifically,
in the region near the upper wall, much more uncertainties are presented in the posterior ensemble
when a non-informative, constant variance field is used.

\begin{table}[htbp]
  \centering
  \caption{Mesh and computational parameters used in the flow over periodic hills and the flow in a
    square duct.
    \label{tab:paraDA}
  }
    \begin{tabular}[b]{c|c c}
      \hline
      \textbf{cases} & \textbf{periodic hill} & \textbf{square duct} \\
      \hline       \hline
      mesh  ($n_x \times n_y$) & $50 \times 30$  & $30 \times 30$ \\ 
      domain size ($L_x \times L_y \times L_z$) & $9H \times 3.306H  \times 0.1H$ & $0.4D \times 0.5D  \times 0.5D$ \\
      $\Delta x \times \Delta y \times \Delta z$ in $y^+$ &   $35 \times [2,65] \times 850$ &
      $24 \times [1.4,30]  \times [1.4,30]$\\
      first grid point in $y^+$ & $\sim 1$, below $2y^+$ in most region & 0.7 \\
      \hline
      number of samples $N$ & \multicolumn{2}{c}{60} \\
      fields with uncertainty & $\xi$, $\eta$, $k$  & $\xi$, $\eta$\\
      number of modes $m$ per field &  $16$ &  $8$ \\
      length scale$^{(a)}$  & H & 0.1D  \\
      number of observation &  18  & 25$^{(b)}$ \\
      std.~dev.~observation noise ($\sigma_{obs}$) & \multicolumn{2}{c}{10\% of truth}  \\
      \hline        
    \end{tabular} \\
    \flushleft
    (a)  Normalized by hill crest height $H$ and domain size $h$ for the periodic hill case and square duct
    case, respectively. \\
    (b)  Only 13 points of velocity data are supplied effectively due to the diagonal symmetry.
\end{table}

\subsubsection{Results}
\label{sec:pehill-res}

\begin{figure}[htbp]
  \centering
    \subfloat[mode 1]{\includegraphics[width=0.24\linewidth]{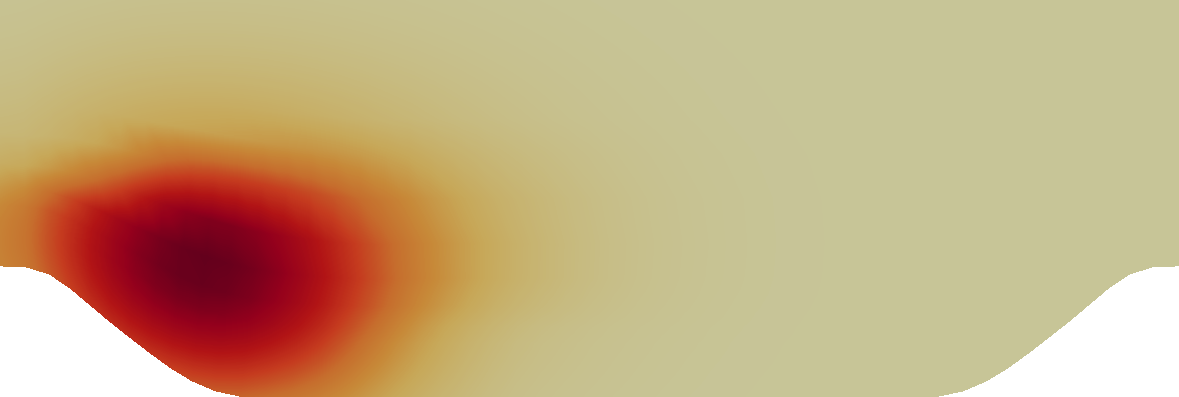}}\hspace{0.1em}
    \subfloat[mode 2]{\includegraphics[width=0.24\linewidth]{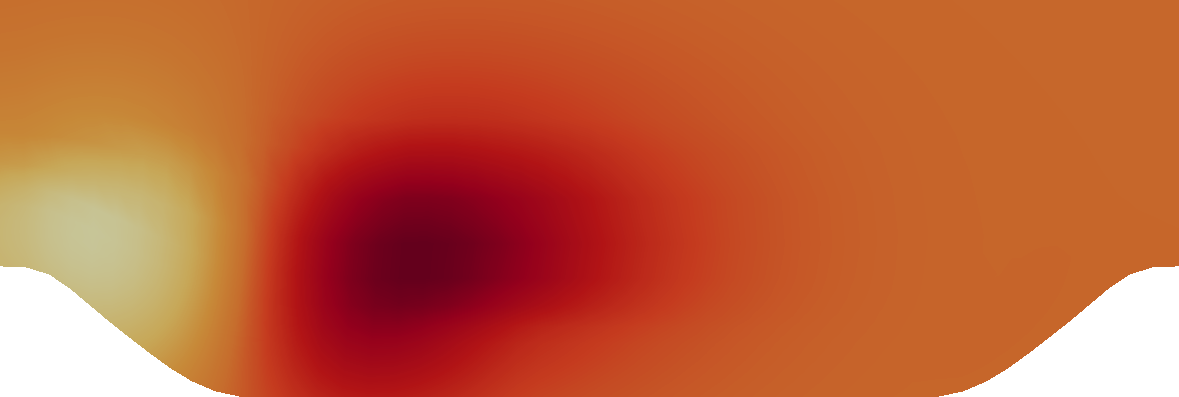}}\hspace{0.1em}
    \subfloat[mode 3]{\includegraphics[width=0.24\linewidth]{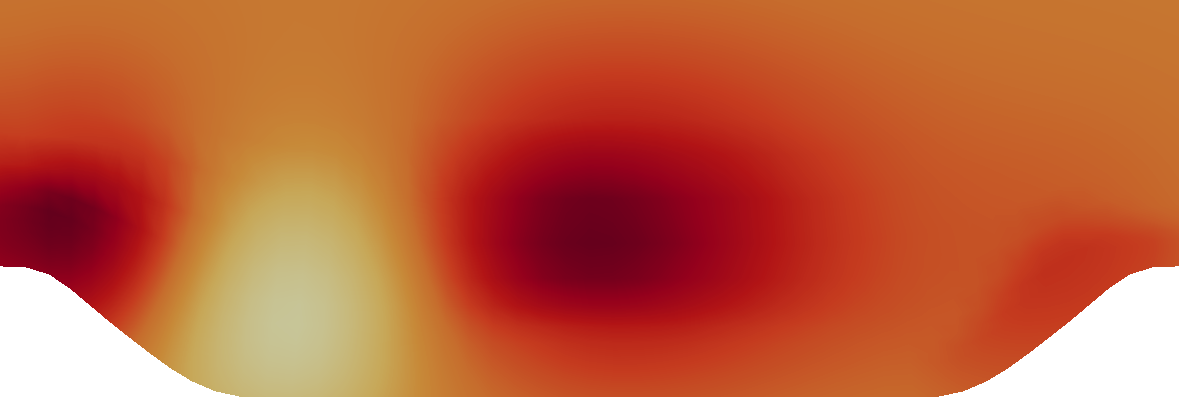}}\hspace{0.1em}
    \subfloat[mode 4]{\includegraphics[width=0.24\linewidth]{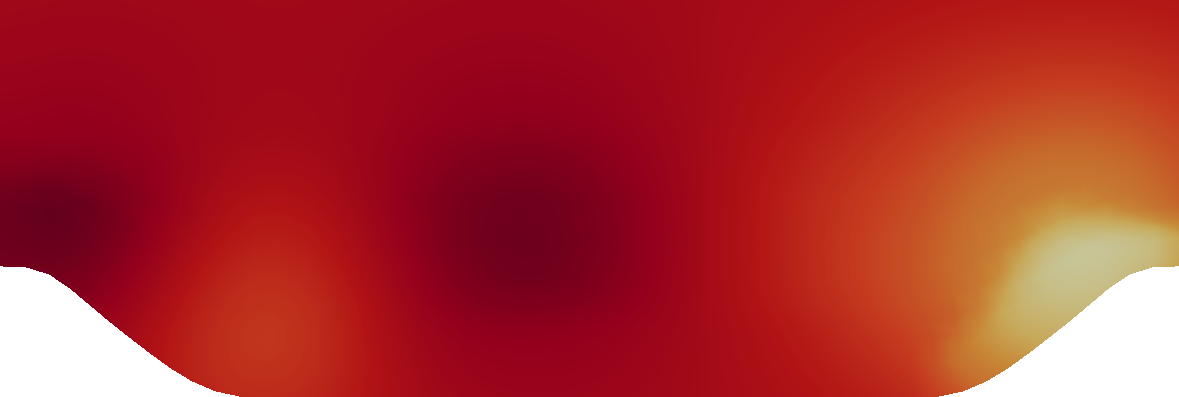}}\\
    \subfloat[mode 5]{\includegraphics[width=0.24\linewidth]{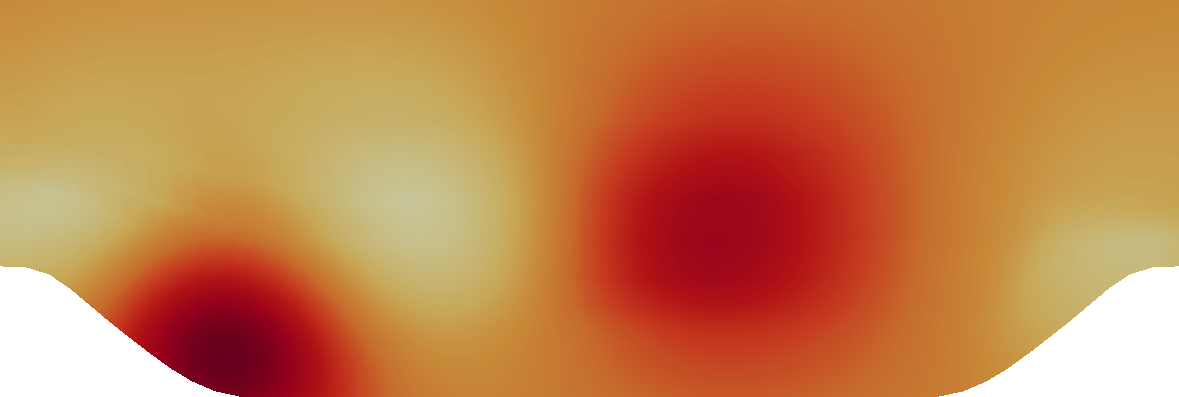}}\hspace{0.1em}
    \subfloat[mode 6]{\includegraphics[width=0.24\linewidth]{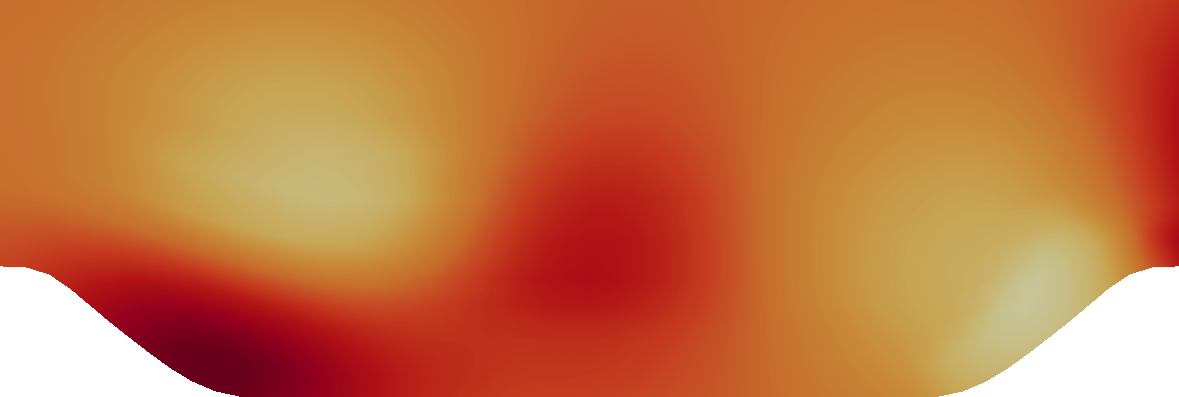}}\hspace{0.1em}
    \subfloat[realization 1]{\includegraphics[width=0.24\linewidth]{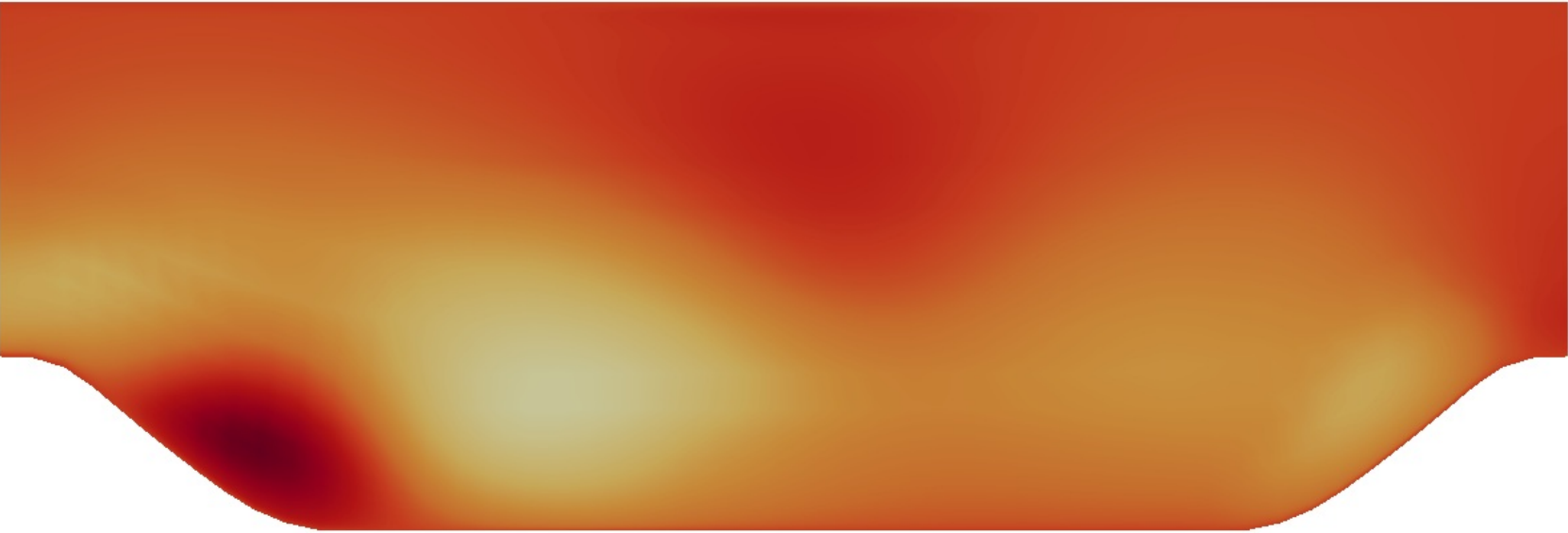}}\hspace{0.1em}
    \subfloat[realization 2]{\includegraphics[width=0.24\linewidth]{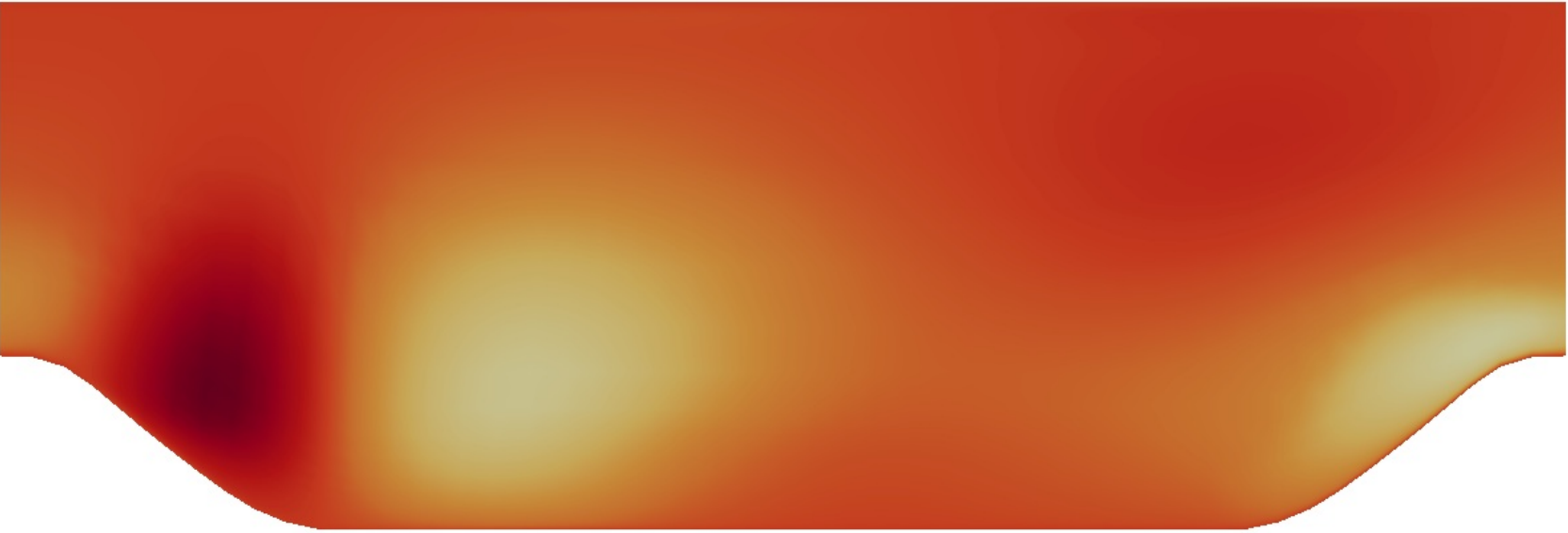}}\\
    \caption{Illutstration of KL expansion modes of the periodic hill case. All the modes have been
      shifted and scaled into the range between 0 (lightest) and 1 (darkest) to facilitate
      presentation, and the legend is thus omitted. Panels (a) to (f) represent modes 1 to 6,
      respectively. Lower modes are more important.  Panels (g) and (h) show the turbulent kinetic
      energy associated with two typical realizations of the Reynolds stress discrepancy fields.  }
  \label{fig:modes_pehill}
\end{figure}

The first six modes of the KL expansion are presented in Fig.~\ref{fig:modes_pehill} along with two
typical realizations. This is to illustrate the uncertainty space of the Reynolds stress discrepancy
field (or more precisely its projections $\delta^\xi$, $\delta^\eta$ and $\delta^k$) . All the modes
have been shifted and normalized to the range $[0, 1]$. It can be seen that in all the models and
the realizations the variations mostly concentrate in the three pre-specified regions (recirculation
zone, free shear region, and the reattached flow windward of the hill), and the upper part of the
channel has rather small variations. This is consistent with our physical prior knowledge specified
through the variance field design.

Accurate predictions of the recirculation and the reattachment of the flow are of the most interest
in the flow over periodic hills. Therefore, we identify three quantifies of interest for this case:
(1) the velocity field, in particular the velocities in the recirculation zone and reattached flow
region windward of the hill, (2) the distribution of shear stresses $\tau_w$ on the bottom wall, and
(3) the reattachment point $x_{attach}$. Other quantities that are important in engineering design
and analysis (e.g., friction drag, form drag, size of separation bubble) are closely related to the
three QoIs above.

\begin{figure}[!htbp]
  \centering
   \hspace{2em}\includegraphics[width=0.5\textwidth]{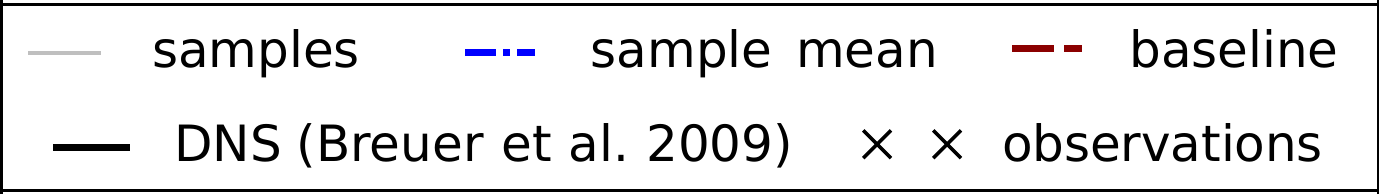}
   \subfloat[Prior velocities ensemble]{\includegraphics[width=0.75\textwidth]{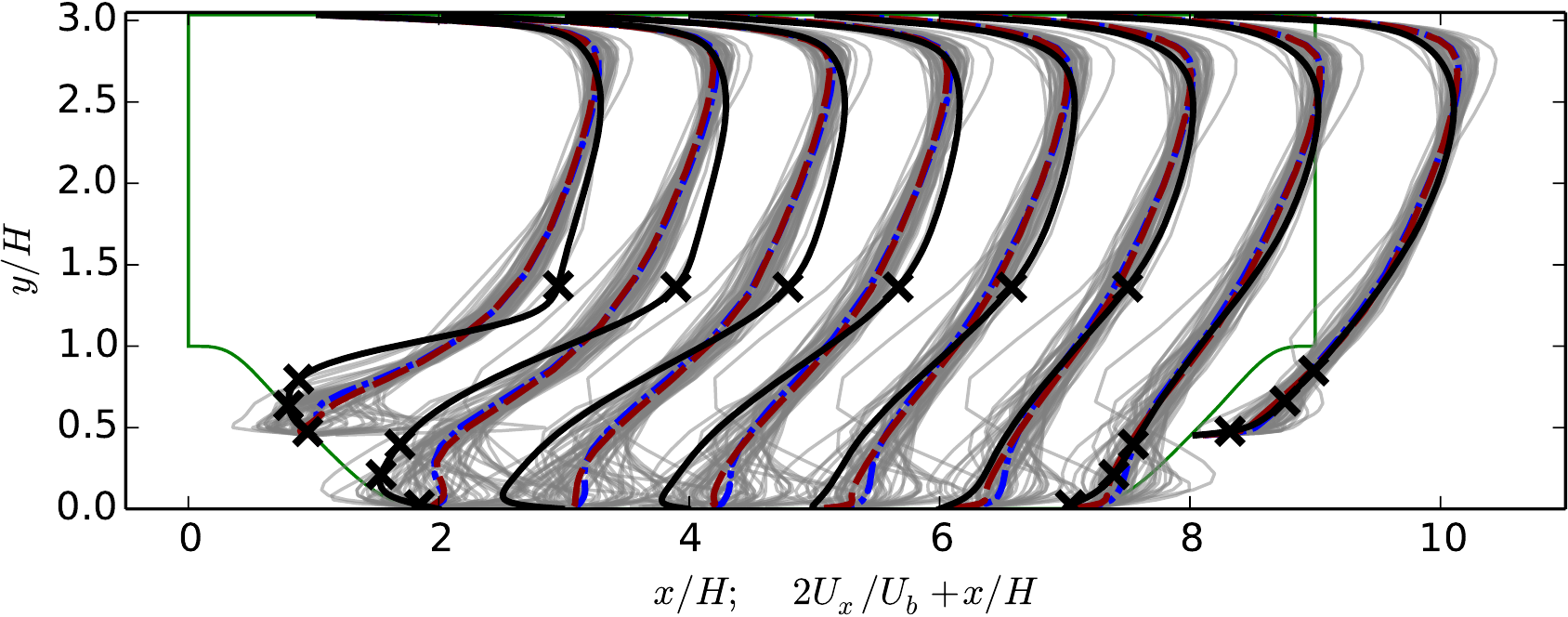}}\\
   \subfloat[Posterior velocities ensemble]{\includegraphics[width=0.75\textwidth]{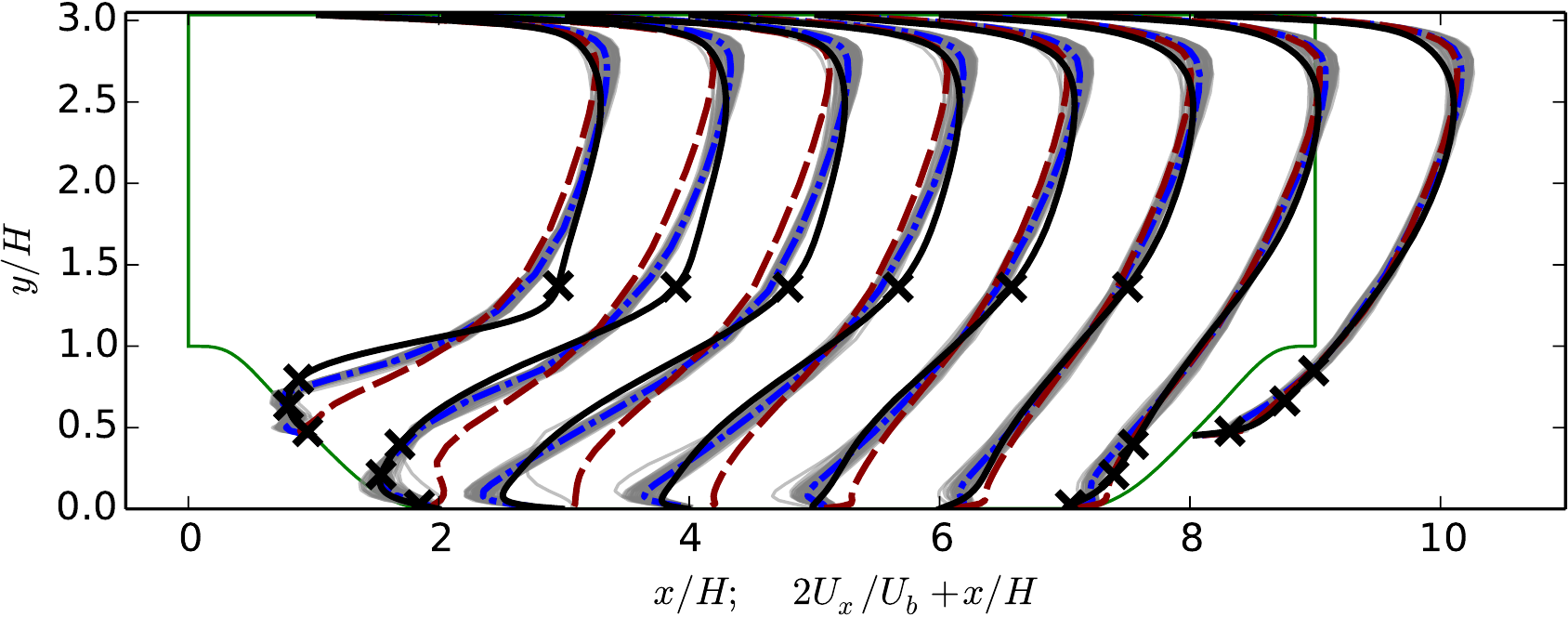}}\\
   \caption{The prior and posterior ensembles of velocity profile for the flow over periodic hill at
     eight locations $x/H = 1, \cdots, 8$ compared with benchmark data and baseline results. The
     locations where velocities are observed are indicated with crosses ($\times$).}
  \label{fig:U_comp_pehill}
\end{figure}

The prior and posterior ensembles of the velocities are presented in Fig.~\ref{fig:U_comp_pehill}
with comparison to the DNS benchmark results. The geometry of the domain is also shown to facilitate
visualization.  From Fig.~\ref{fig:U_comp_pehill}a it can be seen that the prior mean velocity
profiles are very close to those from the baseline RANS simulation, with only minor differences at a
few locations (e.g., near the bottom wall at $x/H =4$, 5, and 6). This is not surprising, since the
Reynolds stresses prior ensemble use the RANS modeled Reynolds stress $\bstaurans$ as the mean. In
other words, the ensemble is obtained by introducing perturbations to the $\bstaurans$. Therefore,
the similarity between the velocity profiles in the baseline simulation and those of the prior
ensemble indicates that the mapping from Reynolds stress to velocity is approximately linear with
respect to the perturbations introduced to the prior Reynolds stresses ensemble. Clearly, both the
baseline velocities and the prior mean velocities deviate significantly from the benchmark results,
particularly in the recirculation region (leeward of the hill).  From Fig.~\ref{fig:U_comp_pehill}
it can be seen that the posterior ensemble mean of the velocities along all the lines are
significantly improved compared to the baseline results.

The remaining differences between the obtained posterior mean and the benchmark data can be
  attributed to two sources: (1) the sparseness of the observation data, and (2) the inadequacy of
  the proposed inference model, specifically, the posterior mean Reynolds stress does not reside in
  the space $\mathcal{A}$ spanned by the prior ensemble. However, note that obtaining the correct
  Reynolds stresses is a sufficient but not necessary condition to infer the correct velocities.
For example, if the divergence of the true Reynolds stresses resides in the space spanned by
  the prior ensemble, the true mean velocity can still be obtained.  This is not surprising since it
  is the divergence of the Reynolds stress tensor field that appears as source term in the RANS
  momentum equation. To illustrate this point, it is particularly interesting to
  investigate the scenario when large amounts of data are available, since any remaining
  discrepancies should then be explained solely by the inadequacy of the inference processes.  We
  performed an experiment where all the benchmark velocities along ten sampled lines at $x/H= 0,
  0.5, 1, 2, \cdots, 8$ were used as observations.  In this scenario the posterior mean velocities
  agree with the benchmark data very well, even in the regions between the sample lines, where no
  data are available.  However, significant discrepancies still remain between the inferred
  posterior mean of the Reynolds stresses and the benchmark.  We argue that the inability to obtain
  the correct Reynolds stress field is \emph{not an intrinsic limitation} of the proposed method.
  Rather, it can be explained by the non-unique mapping between Reynolds stress and velocities as
  described by the RANS equations. That is, two distinctly different Reynolds stress fields can lead
  to identical or very similar velocity fields, because the divergence of the Reynolds stress
  appears in the RANS equation as pointed out above.  The non-unique mapping is further discussed in
  Section~\ref{sec:success}.  However, when we assume that some sparse measurements of Reynolds
  stresses are available, which admittedly are difficult to obtain in practical experiments, the
  inferred Reynolds stresses did improve significantly. The results are omitted here for brevity.

In order to obtain the true Reynolds stresses, the posterior mean Reynolds stress must reside
  in the space $\mathcal{A}$ spanned by the prior ensemble, but in practical inferences there is no
  guarantee this will be the case. There are two reasons for this. First, uncertainties are only
  introduced to the magnitude ($k$) and shape ($\xi$ and $\eta$) of the baseline Reynolds stresses,
  and not to the orientations ($\mathbf{v}_1$, $\mathbf{v}_2$, and $\mathbf{v}_3$). Second, a
  limited number of modes are retained in the KL expansion, which correspond to very smooth fields
  of Reynolds stress discrepancies. Therefore, if we think of the true Reynolds stress as residing
  in a high-dimensional space, in the current framework we assume that the truth is reasonably close
  to the baseline prediction $\bstaurans$, and thus we only search the vicinity of $\bstaurans$ for
  realizable candidates. This is justified by the confidence that the chosen baseline RANS model is
  rather capable, usually backed by previous experiences accumulated by the community on the model
  of concern.

  Finally, we emphasize that if the true Reynolds stresses do reside in the space spanned by the
  prior ensemble, the posterior mean velocity and the Reynolds stresses would indeed coincide with
  the truths. This scenario could occur if the baseline Reynolds stress $\bstaurans$ only differs
  from the true Reynolds stress in magnitude $k$ and shape $\xi$ and $\eta$, and the discrepancy is
  smooth enough to be represented by the chosen number of modes. However, both scenarios are rather
  unlikely in any nontrivial cases.  For verification purposes we have designed a case of flow over
  periodic hills with synthetic data (as opposed to DNS data) that satisfies the requirements above,
  and have confirmed that the obtained posterior mean velocity indeed exactly agrees with the truth
  in this case, and that the true Reynolds stress can also be obtained. The results are detailed in
  a separate work~\cite{mfu2}.

This claim that the prior has small influence to the prior is apparently contradictory to the
  Bayesian inference theory, which states that the posterior is proportional to the product of the
  prior and the likelihood informed by the observation data. However, in the ensemble Kalman
  method used in this work, the observation data are imposed on the prior iteratively (albeit with
  different noises in each iteration). As a result, the influence of the prior on the posterior
  diminishes as the posterior proceeds to statistical convergence.  From a practical perspective,
  the iterative ensemble Kalman method can also be interpreted as an optimization procedure (e.g.,
  that used by Parish et al.~\cite{parish2016paradigm}), with the prior corresponding to the
  initial guess. This interpretation, as an alternative to the Bayesian interpretation, has been
  advocated in the literature~\cite{iglesias2013ensemble,schillings2016analysis}. 

\begin{figure}[!htbp]
\centering
\hspace{2.5em}\includegraphics[width=0.55\textwidth]{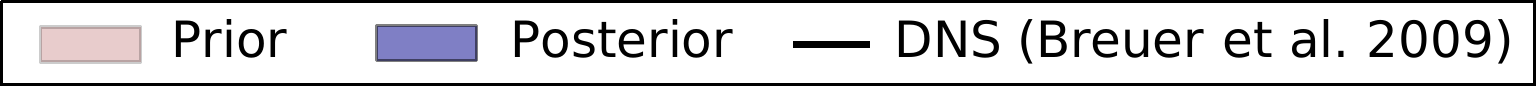}\\
\includegraphics[width=0.9\textwidth]{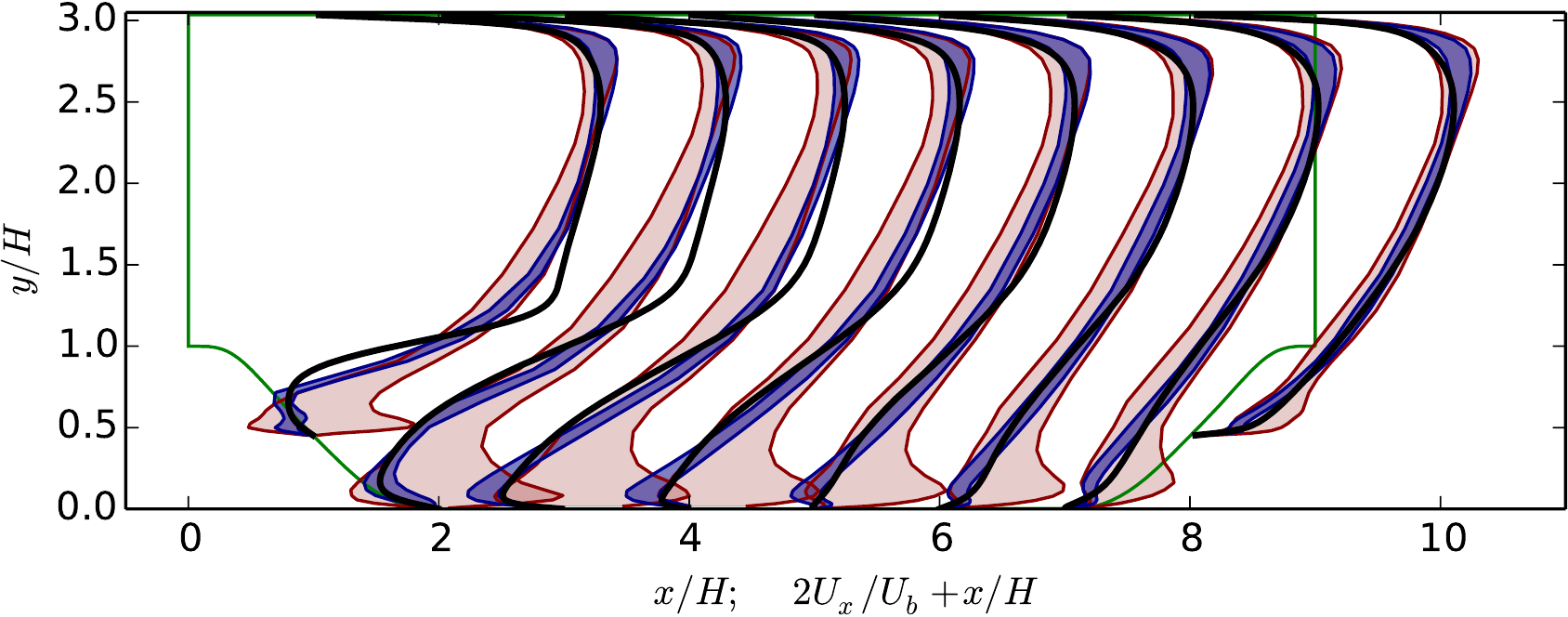}
\caption{The 95\% credible intervals of the prior (light/pink shaded region) and posterior
  (dark/blue shaded region) ensembles of velocity profiles for the flow over periodic hills.}
\label{fig:Ux_pehill_confidence}
\end{figure}

Figure~\ref{fig:Ux_pehill_confidence} shows the 95\% credible intervals estimated from the data in the
prior and posterior velocity ensembles. That is, at each point, 95\% of the samples fall within the
shaded region (light/pink shaded for the prior and dark/blue shaded for the posterior).  Note that
the credible intervals shown here are point estimations and do not contain information on the
spatial correlations of the velocity profiles.  It can be seen from
Fig.~\ref{fig:Ux_pehill_confidence} that the 95\% credible interval in the posterior is
significantly narrowed compared to that in the prior, which suggests that the model form uncertainty
is reduced by incorporating the velocity observation data.  Such a reduction of uncertainty is more
visible in the recirculation zone, where more observation data are available. In contrast, the prior
uncertainty near the upper wall largely remains, which is due to the lack of observation data in
this region.

\begin{figure}[!htbp]
  \centering
   \includegraphics[width=0.75\textwidth]{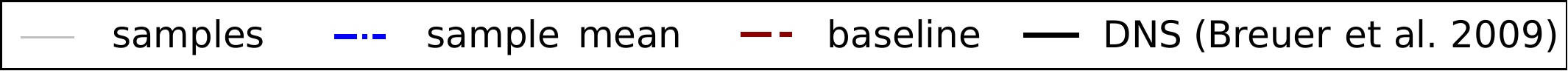}
   \subfloat[Prior ensemble]{\includegraphics[width=0.5\textwidth]{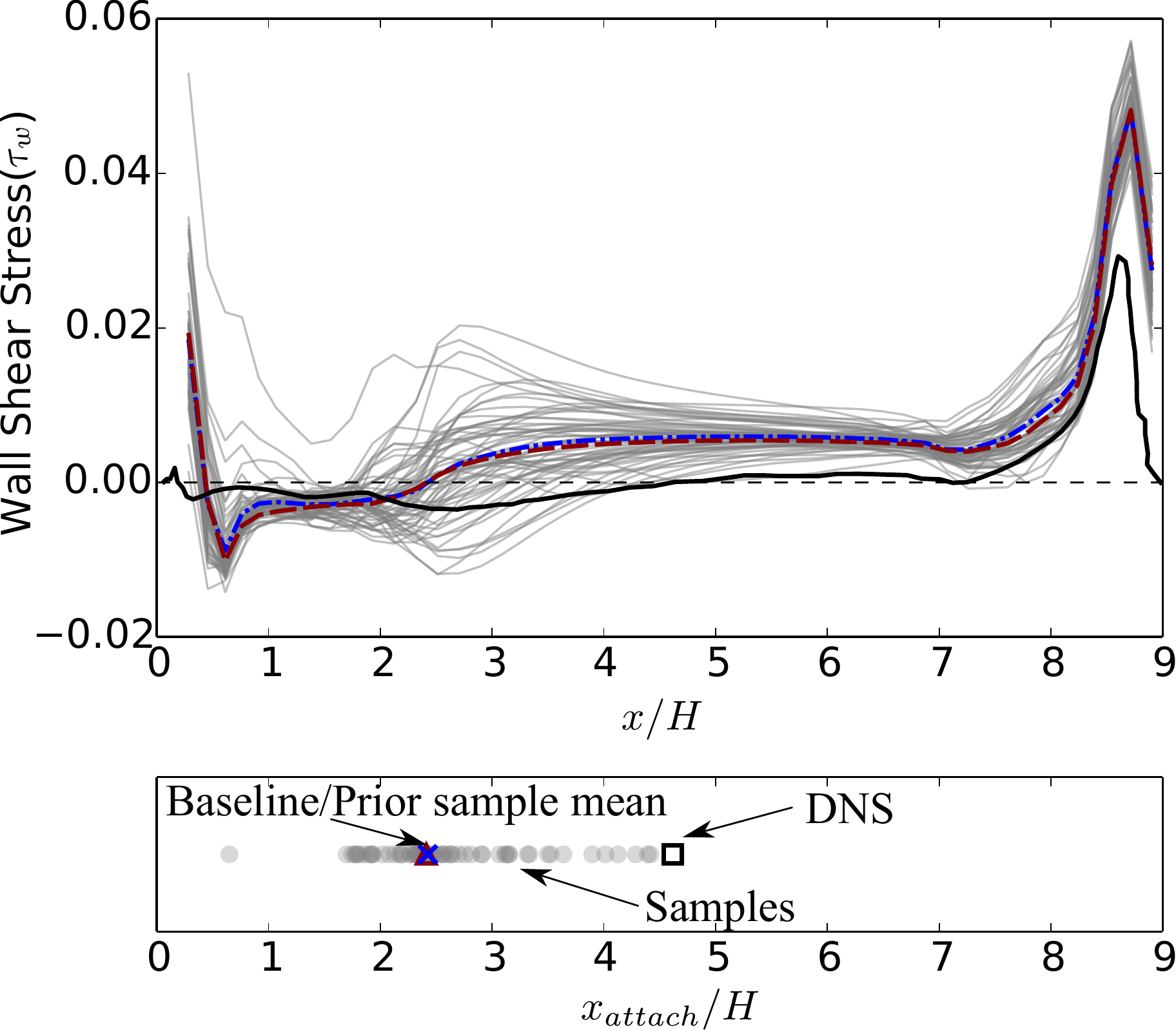}}
   \subfloat[Posterior ensemble]{\includegraphics[width=0.5\textwidth]{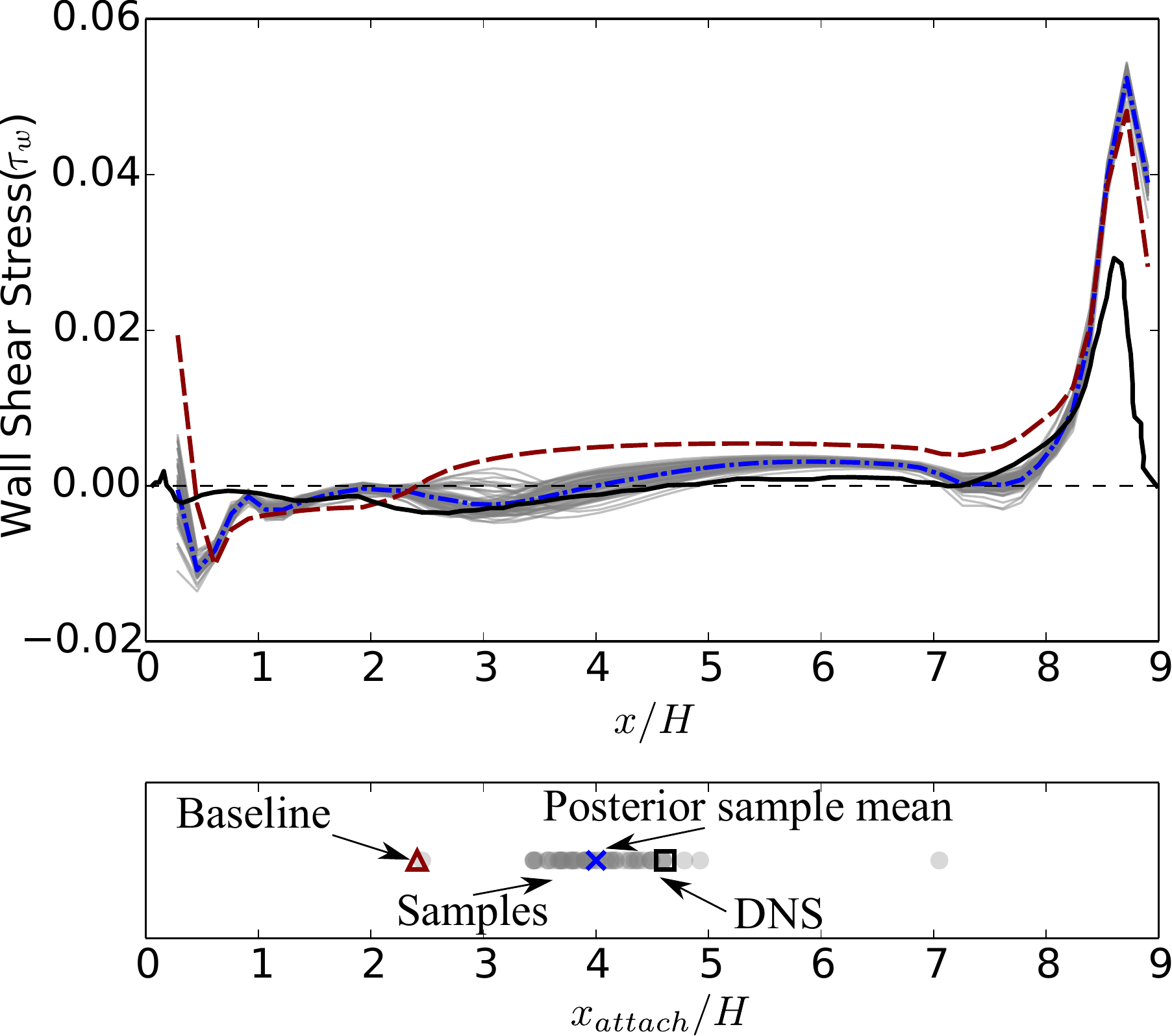}}\\
   \caption{(a) Prior ensemble and (b) posterior ensemble of the bottom wall shear stress $\tau_w$
     (\textit{top panels}) and reattachment point $x_{attach}$ (\textit{bottom panels}) for the flow
     over periodic hills. The region with negative shear stress $\tau_w$ indicates the extent of
     recirculation zone on the bottom wall.  The reattachment point is the downstream end of the
     recirculation zone, which can be determined by the location at which the wall shear stress
     changes from negative to positive.  Note that certain samples in the ensemble have two
     recirculation zones that are very close to each other. In these cases the reattachment point of
     the downstream one is taken.}
  \label{fig:tau_wall}
\end{figure}

The other two QoIs, bottom wall shear stress $\tau_w$ and the reattachment point $x_{attach}$, are
shown in Fig.~\ref{fig:tau_wall}. Similar to the velocity profiles in Fig.~\ref{fig:U_comp_pehill},
both prior and posterior ensembles are presented and compared with benchmark data and baseline
results.  It can be seen from Fig.~\ref{fig:tau_wall}a that the prior ensemble means of both
$\tau_w$ and $x_{attach}$ deviate from the benchmark DNS data significantly.  In particular, the
baseline RANS simulation predicts a much smaller recirculation zone than the
truth. Figure~\ref{fig:tau_wall}b shows that in most of the region (between $x/H=1$ and
$x/H=8$) the posterior ensemble mean has better agreement with the DNS data than the baseline
results. In fact, in this region, all samples in the posterior ensemble have better agreement with the
benchmark than the baseline results in terms of both wall shear stress and reattachment point.  This
improvement demonstrates the merits of the current framework. Incorporating observation data and
physical prior knowledge indeed leads to improved predictions of both QoIs.

It is noted that in the immediate vicinity of the hill crest, i.e., near $x/H = 0.5$ and $x/H =8.5$,
the posterior ensemble is similar to or even slightly deteriorated compared to the baseline in terms
of agreement with DNS data. The reason is that in this region the flow has rapid spatial
variations. Specifically, there is a separation between $0 < x/H < 1$, and a large mean flow
curvature with strong pressure gradient between $8 < x/H < 9$. Consequently, the length scales of
the coherent structures in this region are small, and thus the correlations between this part of the
flow and other regions are weak. On the other hand, no velocity observations are available in this
region.  Here we point out an important fact that the Bayesian inference based on ensemble Kalman
method primarily relies on the \emph{correlation} between the predicted system state variables at
different locations to make corrections. Specifically, the observations only bring information to
the states at the locations correlated to the observed states. Hence, poor prediction is expected
for a region that has neither observations within it nor statistically significant correlations with
the regions that have observations. The role of correlation in the current framework is further
discussed in Section~\ref{sec:correlation}.

The prior and posterior distributions of the reattachment point represented by samples are shown in
Fig.~\ref{fig:tau_wall} (bottom panels).  It can be seen that the bias in the prior distribution as
compared to the DNS data is large, while in the posterior distribution the bias is significantly
corrected.  Moreover, the prior sample scattering is wide, indicating large uncertainties.  In
contrast, the posterior distribution is significantly narrowed, which indicates increased confidence
in the prediction by incorporating velocity observations into the Bayesian inference.

\begin{figure}[!htbp]
\centering
\includegraphics[width=0.55\textwidth]{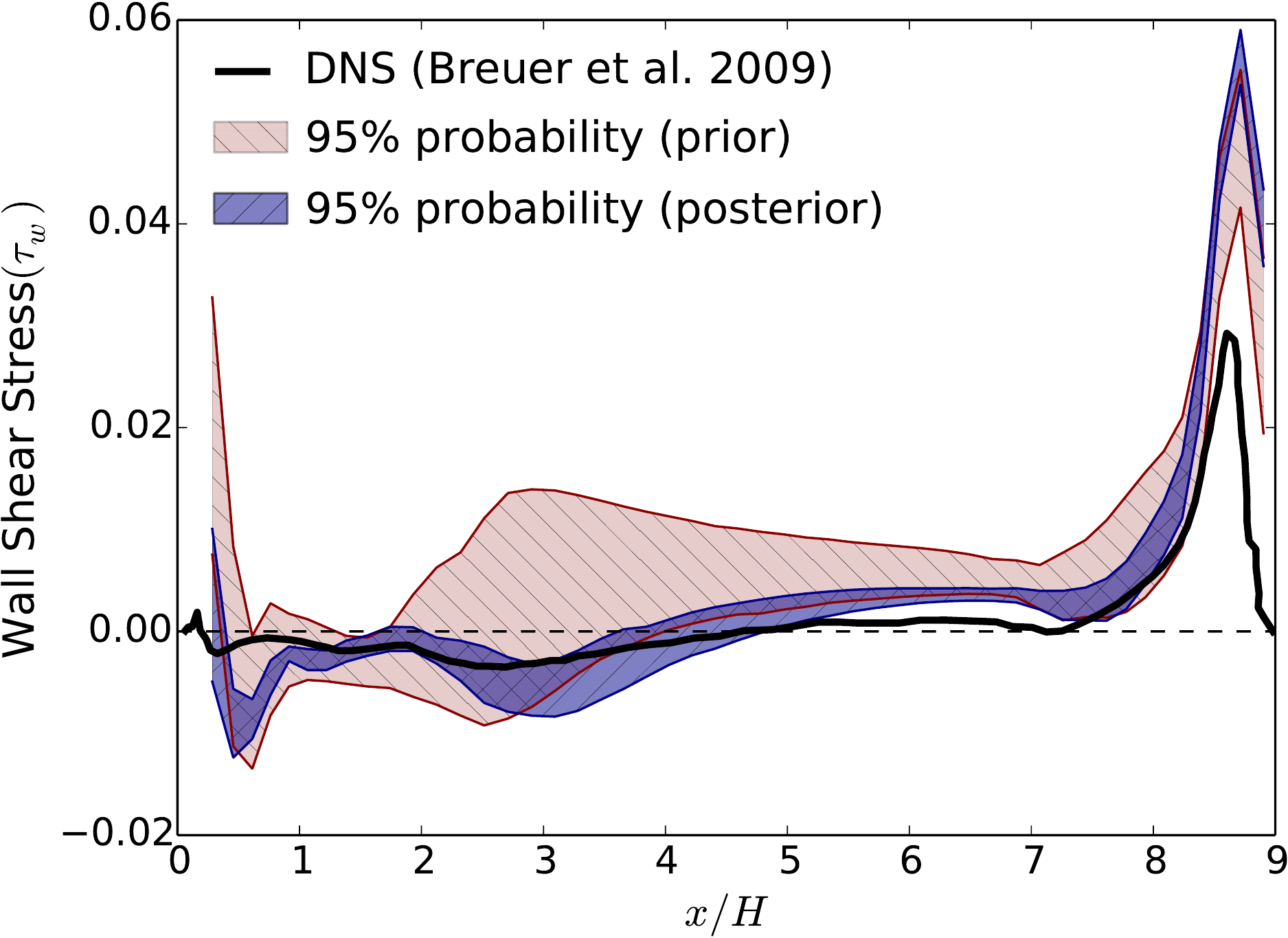}
\caption{ The 95\% credible intervals of the prior (light/pink shaded region) and posterior
  (dark/blue shaded region) ensembles of bottom wall shear stress for the flow over periodic hills.}
\label{fig:shear_confidence}
\end{figure}

The comparison of 95\% credible interval obtained from the prior and posterior ensemble of wall
shear stresses are presented in Figure~\ref{fig:shear_confidence}.  Similar reduction of model-form
uncertainty as shown in Fig.~\ref{fig:Ux_pehill_confidence} is observed here.  Compared to that in
the prior, the 95\% credible interval in the posterior has a much smaller uncertainty and a better
coverage of the benchmark data in the region between $x/H=1$ and 5. This is because there are more
observations available in the vicinity. Admittedly, in some regions, e.g., between $0 < x/H < 1$ and
$8 < x/H < 9$, the posterior credible interval does not improve or even deteriorate compared to the
prior, which is due to the lack of observation data and the relatively small length scale in these
regions as discussed above.  It is noted that in some regions the 95\% credit intervals,
  e.g., in Figs.~\ref{fig:Ux_pehill_confidence} and~\ref{fig:shear_confidence}, failed to cover the
  truth, which indicates that the current method should still be used with caution when making
  high-consequence decisions. The iterative ensemble Kalman method tends to underestimate
  uncertainties in the posterior distributions, a difficulty shared by many other maximum likelihood
  estimators as well~\cite{neuman2003maximum}.

\begin{figure}[!htbp]
 \centering
  \hspace{2em}\includegraphics[width=0.75\textwidth]{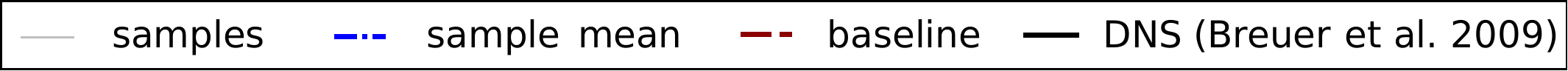}
  \subfloat{\includegraphics[width=0.9\textwidth]{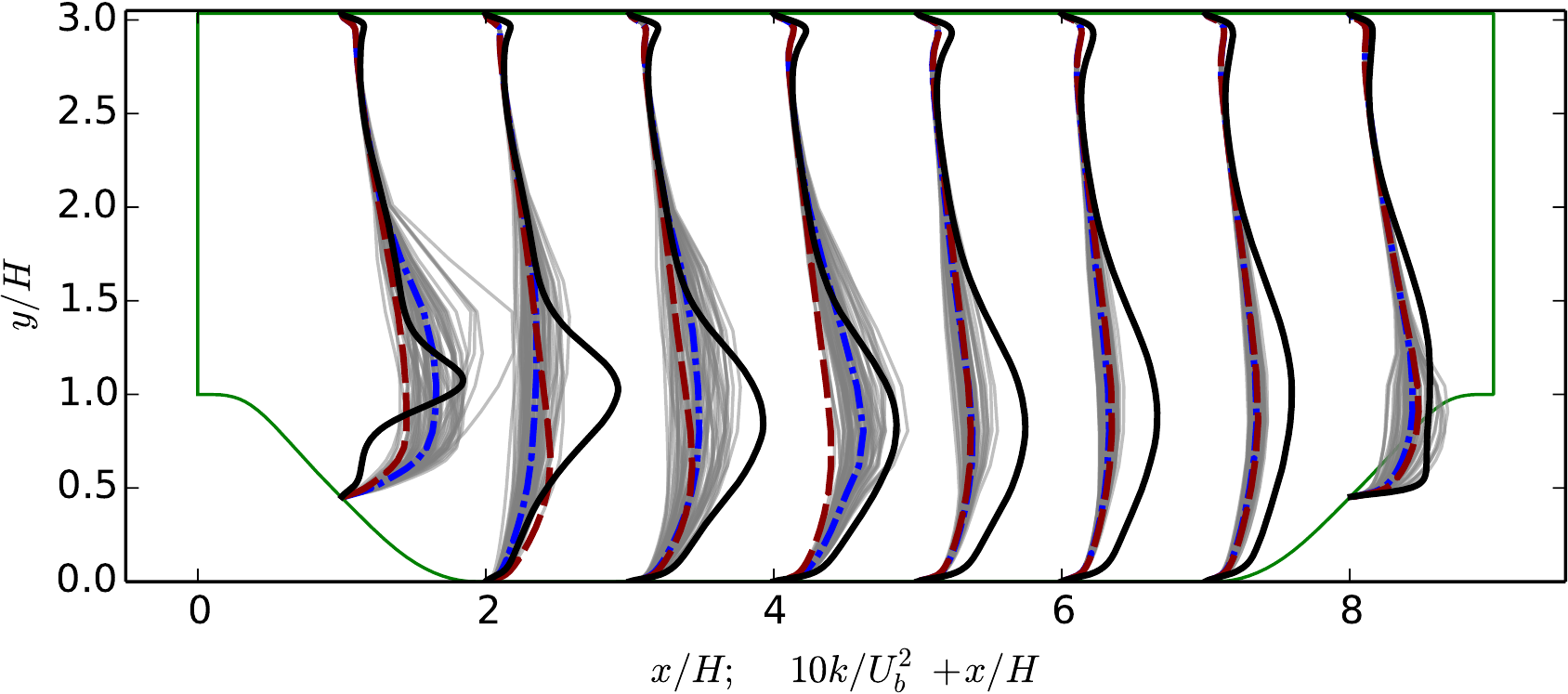}}\\
  \caption{Posterior ensemble of the turbulent kinetic energy $k$ with its mean compared to the
    baseline results and the benchmark DNS data.  The prior ensemble is omitted for $k$, since its
    mean is the same as the baseline prediction.}
 \label{fig:tke_pehill}
\end{figure}

Figure~\ref{fig:tke_pehill} shows that the bias in the turbulent kinetic energy (TKE) from baseline
RANS prediction has been partly corrected, especially for the upstream region. It is possible that
the production of TKE due to the instability in the free-shear region after the separation is the
driving factor. Consequently, the improved prediction of TKE in this region leads to the corrections
for the velocities and other QoIs in the entire field. However, note that the posterior mean of TKE
is not necessarily better than the baseline results at all locations.  The TKE levels immediately
downstream of the hill crest have been increased, but at the downstream locations the posterior mean
are not significantly better than the baseline.  In the process of minimizing misfit with the
observations, some compromises are inevitably made, with some regions such as the upstream
experiencing more corrections than other regions such as the downstream region.  A possible
explanation is that the TKE in the free-shear region has stronger correlations with the velocities
at the observed locations. However, it is also possible that the TKE does not necessarily need
  to be improved to provide better velocity field due to the non-unique mapping from Reynolds stress
  field to velocity field, as has been discussed above. More detailed discussion can be found in
  Section~\ref{sec:success}.

\subsection{Fully Developed Turbulent Flow in a Square Duct}

\subsubsection{Case setup}

\begin{figure}[htbp]
  \centering
   \includegraphics[width=0.8\textwidth]{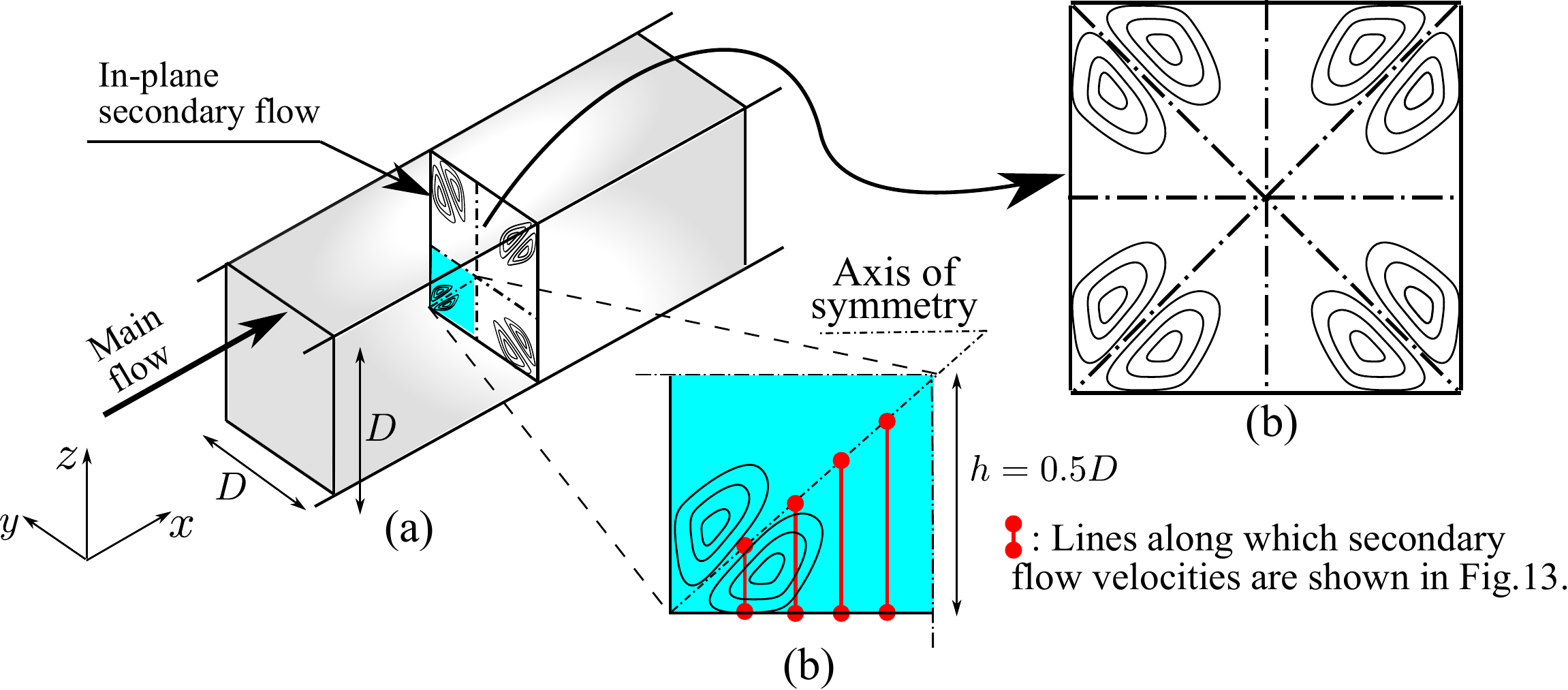}
   \caption{(a) Schematic for the fully developed turbulent flow in a square duct. The $x$ axis is
     aligned with the streamwise direction.  Secondary flows exist in the $y$--$z$ plane, which are
     schematically represented with contours.  (b) Symmetry of the (mean) flow with respect to the
     centerlines in $y$- and $z$-directions and along the diagonals.  (c) The computational domain
     covers only a quarter of the physical domain due to the centerline symmetry. The cross-sections
     along which QoIs (e.g., velocities and Reynolds stress imbalance) are compared to benchmark
     data are also indicated.}
  \label{fig:domain_duct}
\end{figure}

The fully developed turbulent flow in a square duct is a widely known case for which many turbulence
models fail to predict the secondary flow induced by the Reynolds stresses. The geometry of the case
is shown in Fig.~\ref{fig:domain_duct}.  The Reynolds number based on the edge length $D$ of the
square and the bulk velocity $U_b$ is $Re_b=10320$. All lengths presented below are
normalized by the height $h$ of the computational domain, which is half of $D$.  Extensive
benchmark data from DNS are available in the literature~\cite{huser1993direct,agard98}.

Standard computational setup as used in the literature is adopted in this work.  Only one quadrant
of the physical domain is simulated considering the symmetry of the flow with respect to the
centerlines along $y$- and $z$-axes as indicated in Fig.~\ref{fig:domain_duct}. We emphasize here
that our study is concerned with the mean flow, since the objective is to quantify the uncertainties
in RANS simulations.  The instantaneous flows are beyond the scope of our discussions, and they do
not have the symmetries mentioned here.  Non-slip boundary conditions are imposed at the walls and
symmetry boundary conditions (zero in-plane velocities) are applied on the symmetry planes.
Theoretically, one can further reduce the computational domain size to 1/8 of the physical domain by
utilizing the symmetry with respect to the square diagonal. However, this symmetry is not exploited,
as it would be difficult to impose proper boundary conditions on the diagonal.  The symmetry in the
baseline RANS simulation results is implied by the diagonal symmetry of the geometry and boundary
conditions.  When conducting forward RANS simulations with given Reynolds stress fields, caution
must be exercised to ensure that the perturbations introduced to $\bstaurans$ have diagonal
symmetry, which will be discussed later. Otherwise, the posterior velocities may be asymmetric with
respective to the diagonal.


The mesh and computational parameters for this case are shown in Table~\ref{tab:paraDA}. Choice of
parameters can be motivated similarly as in the periodic hill case. A notable difference is that
only uncertainties in the shape of the Reynolds stress (i.e., $\xi$ and $\eta$) are considered in
the square duct flow case. The QoI for this flow is the in-plane flow velocities, which are
primarily driven by the normal stress imbalance $\tau_{yy}-\tau_{zz}$, a quantity that is associated
with the shape of $\bstau$.  The design of the variance field $\sigma$ is based on the same
principle as in the periodic hill flow.  Specifically, we chose $\sigma_{0} = 0.2$ throughout the
field and $\sigma_{local} = 0.5$ at the lower left corner. The length scale of the radial basis
kernel is chosen as 0.1D based on the estimation of length scale of secondary flow (see
Table~\ref{tab:paraDA}). It is known that RANS models have more difficulties in predicting the flow
near the corner, which justifies the large value of $\sigma(x)$ near the corner and the gradual
decrease away from the corner as well as towards the diagonal.  Moreover, the variance field is
chosen to be symmetric along the diagonal of the $y$--$z$ plane in consideration of the flow
symmetry.

\begin{figure}[!htbp]
  \centering
  \includegraphics[width=0.6\textwidth]{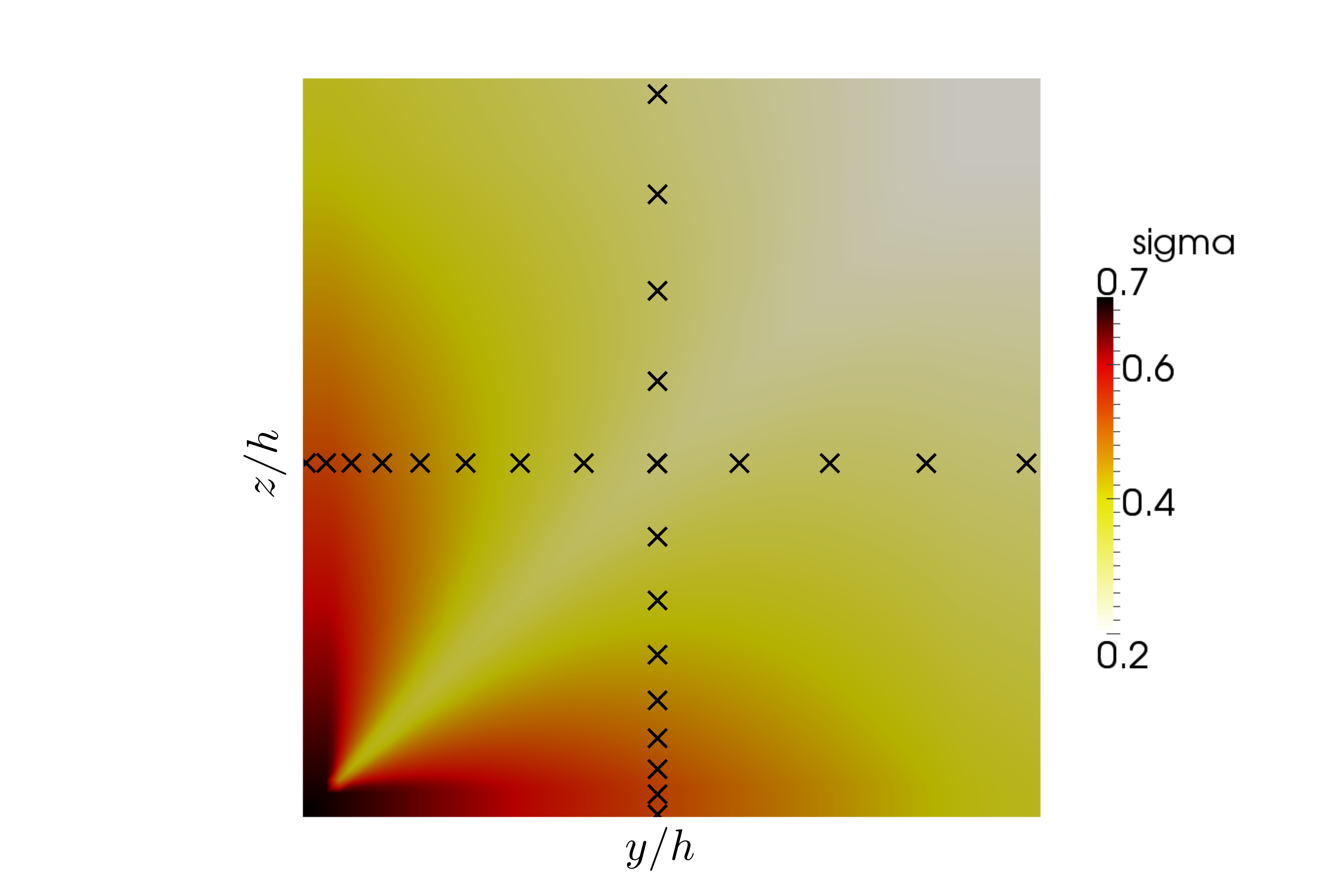}
  \caption{Contour of the variance field $\sigma(x)$ and locations of the observations for the
    square duct flow case. Larger variances are allowed near the corner due to the difficulties RANS
    models have in predicting the secondary flow in this region.  The variance field is chosen to be
    symmetric along the diagonal of the $y$--$z$ plane in consideration of the flow symmetry.}
  \label{fig:sigma_duct_dns}
\end{figure}

\subsubsection{Results}

\begin{figure}[htbp]
  \centering \subfloat[mode1]{\includegraphics[width=0.24\linewidth]{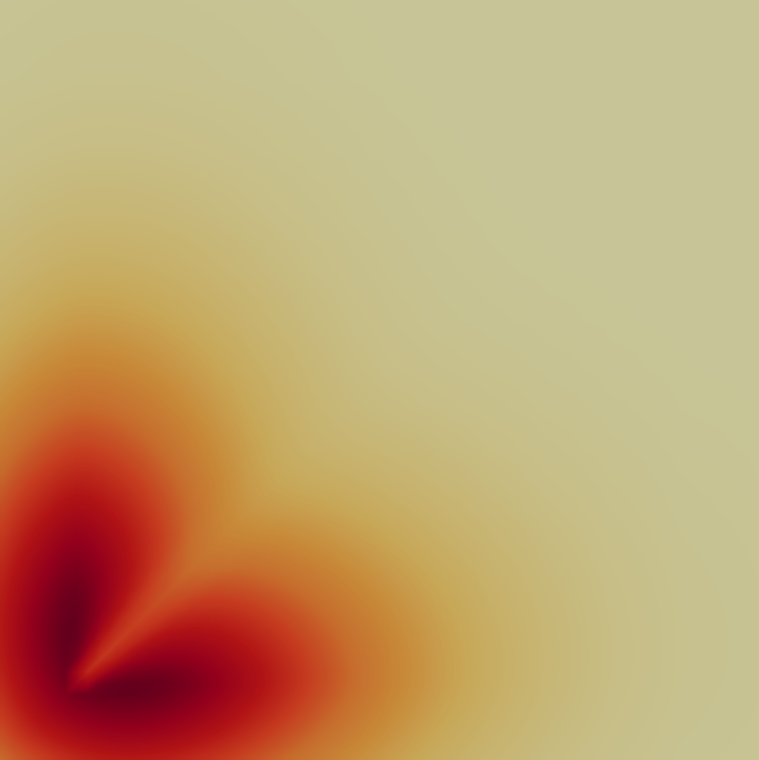}}\hspace{0.02em}
  \subfloat[mode 2]{\includegraphics[width=0.24\linewidth]{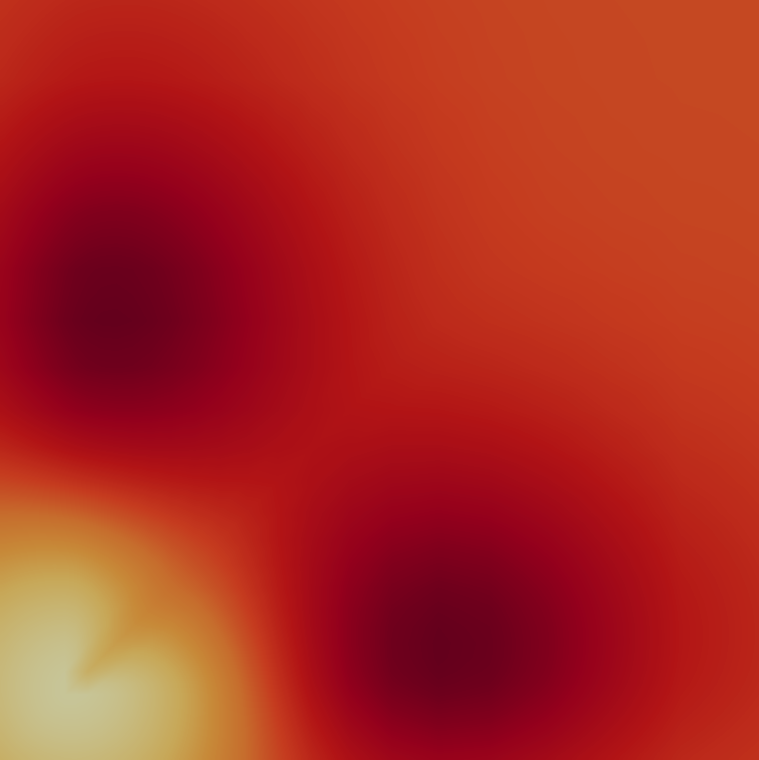}}\hspace{0.02em}
  \subfloat[mode 3]{\includegraphics[width=0.24\linewidth]{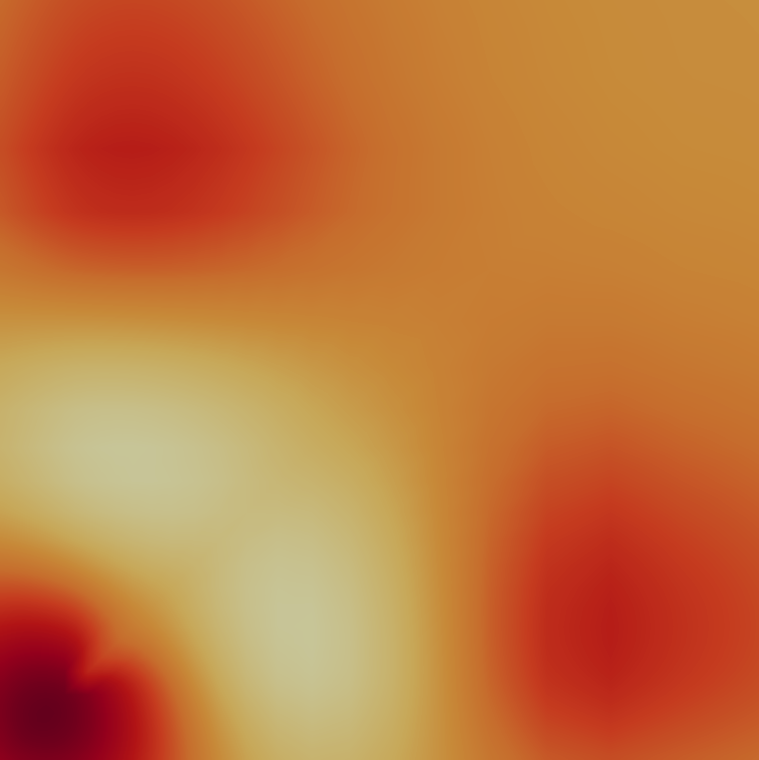}}\hspace{0.02em}
  \subfloat[mode 4]{\includegraphics[width=0.24\linewidth]{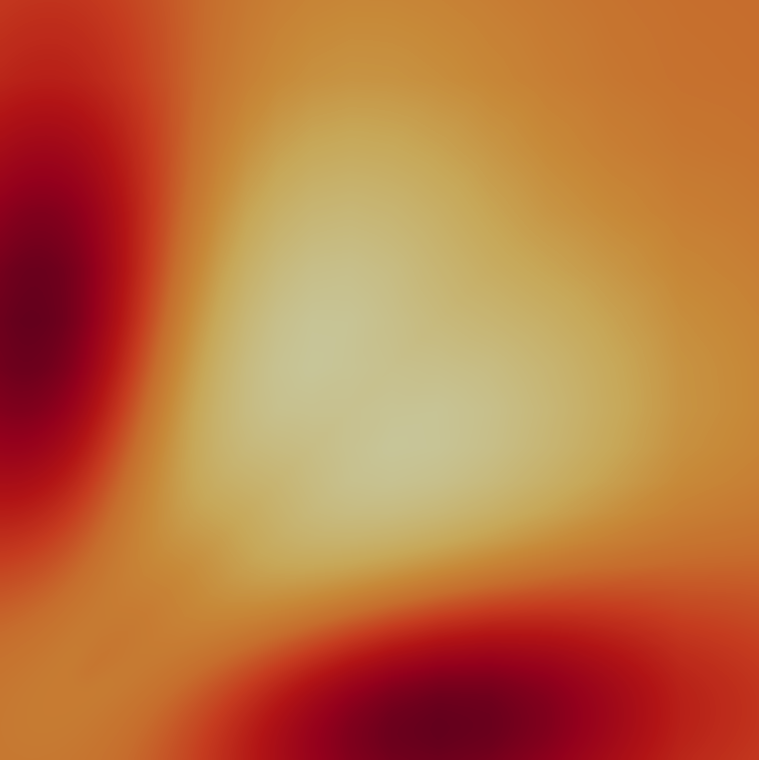}}\\
  \subfloat[mode 5]{\includegraphics[width=0.24\linewidth]{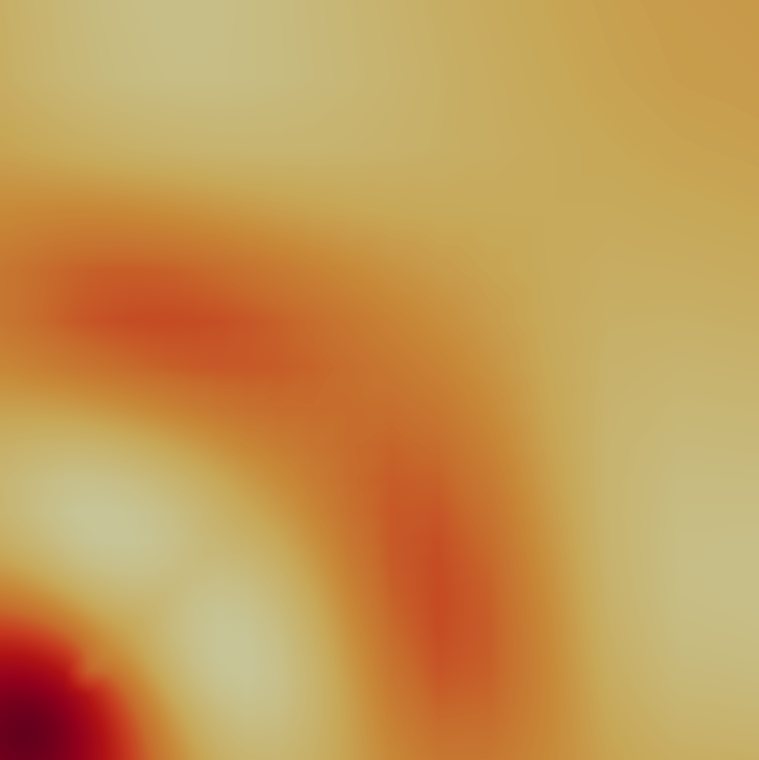}}\hspace{0.02em}
  \subfloat[mode 6]{\includegraphics[width=0.24\linewidth]{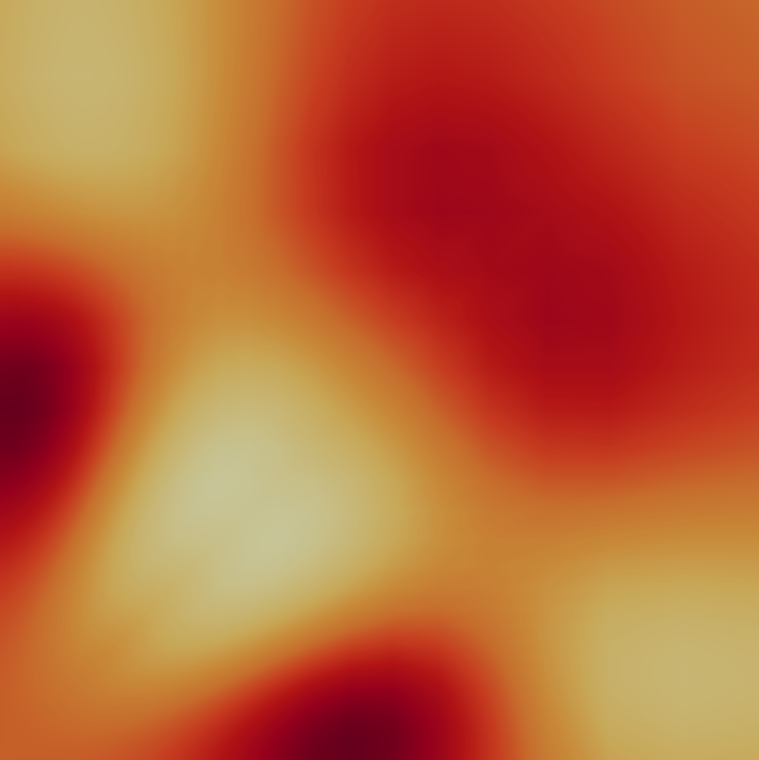}}\hspace{0.02em}
  \subfloat[realization 1]{\includegraphics[width=0.24\linewidth]{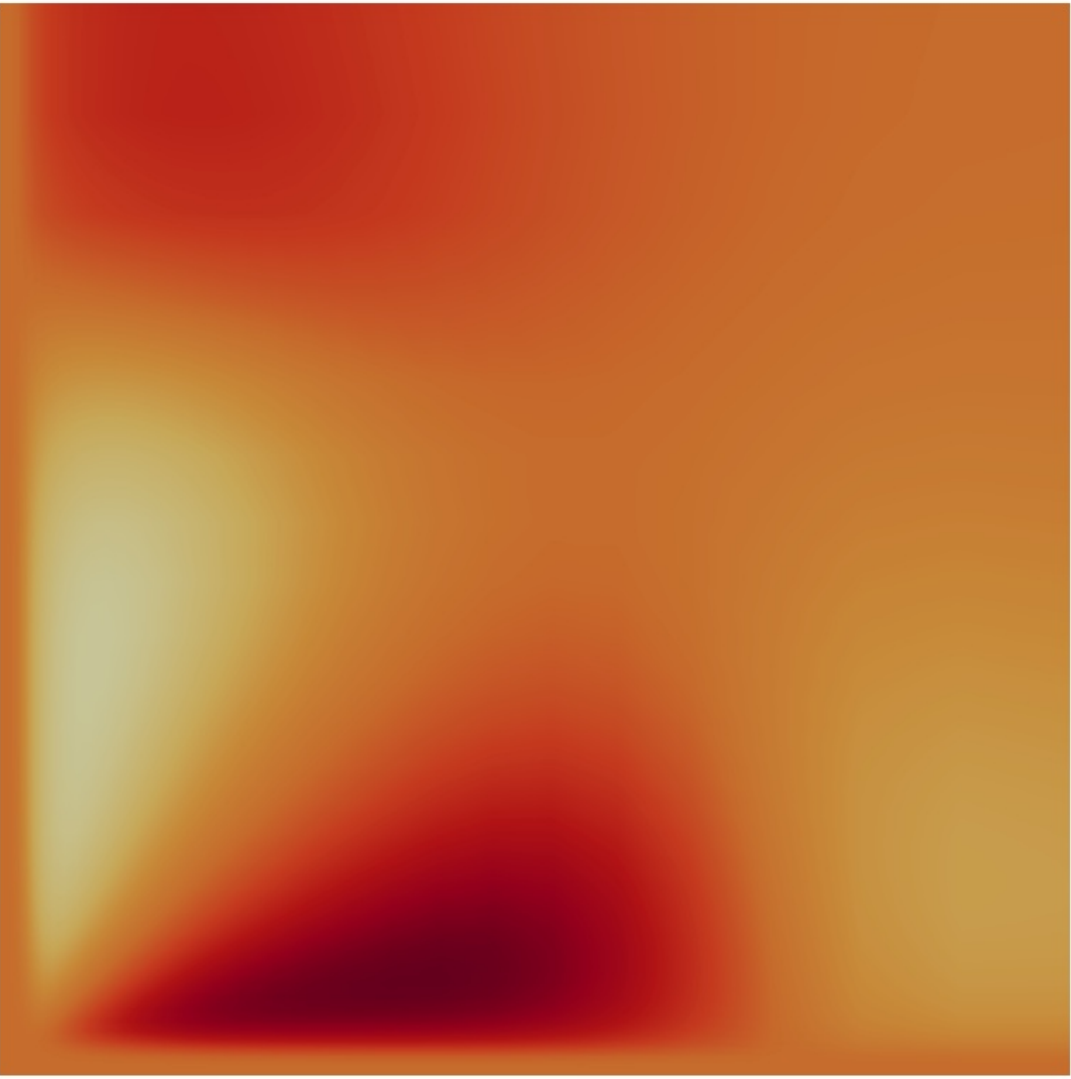}}\hspace{0.02em}
  \subfloat[realization 2]{\includegraphics[width=0.24\linewidth]{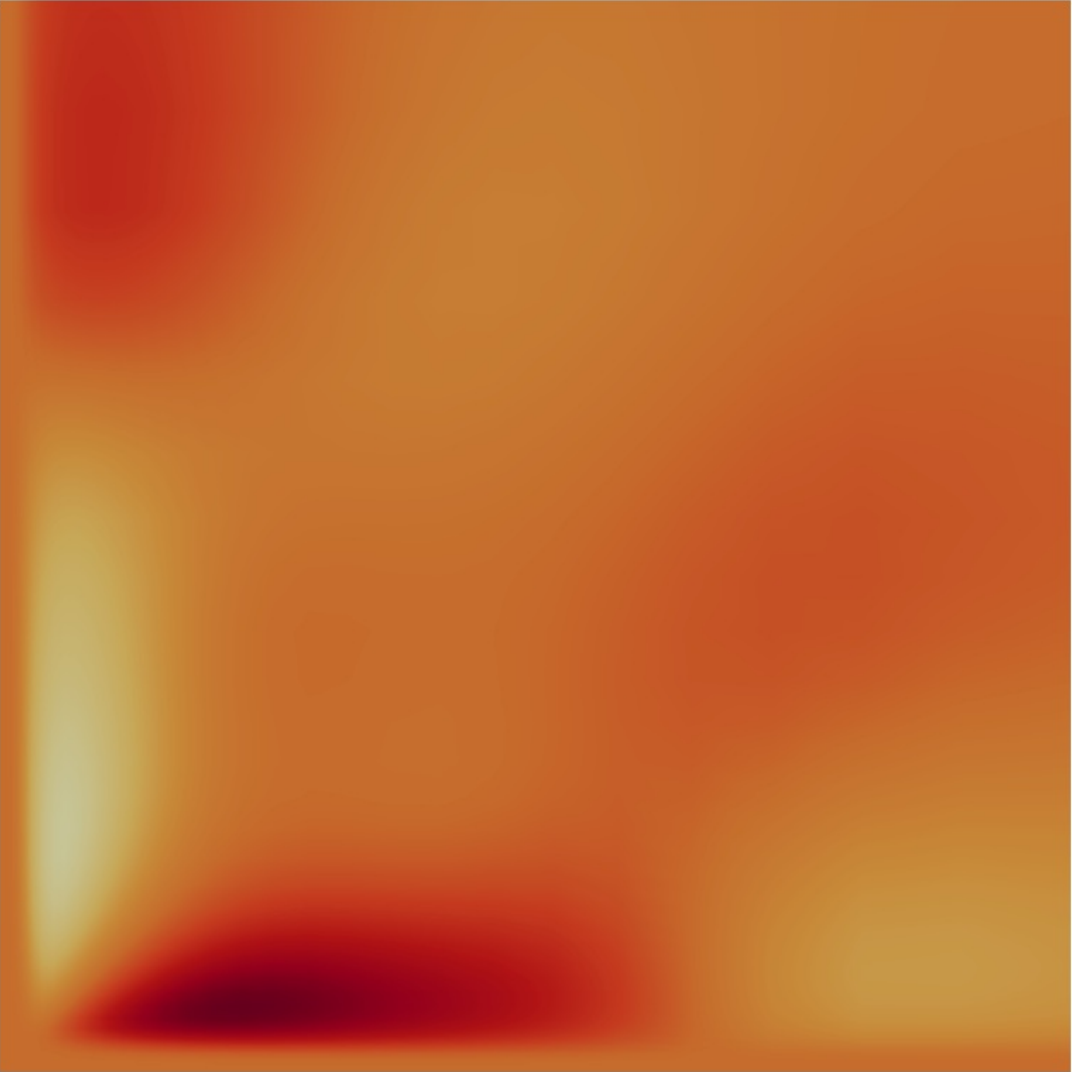}}\\
  \caption{Illustration of KL expansion modes for the square duct flow case. All presented modes
    have been shifted and scaled into a range of 0 (lightest) to 1 (darkest) to facilitate
    presentation, and the legend is thus omitted. Panels (a) to (f) denote modes 1 to 6,
    respectively, with lower modes being more important.  Only the modes with diagonal symmetry are
    retained to guarantee the symmetry of perturbed Reynolds stresses field. Panels (g) and (h) show
    the magnitude of Reynolds stress imbalance $|\tau_{yy}-\tau_{zz}|$ associated with two typical
    realizations of the discrepancy fields.}
  \label{fig:modes_duct}
\end{figure}

The first six modes of KL expansion are shown in Fig.~\ref{fig:modes_duct} along with two typical
realizations. All the modes have been shifted and normalized into the range $[0, 1]$. Only the
diagonally symmetric modes are retained to guarantee the symmetry of the Reynolds stress along the
diagonal, which leads to the symmetry of the posterior velocities.  The observations are obtained
from the DNS data~\cite{huser1993direct} by adding Gaussian random noises as in the periodic hill
flow. Velocities are observed at 25 points as shown in Fig~\ref{fig:domain_duct}, half of which are
distributed along the line $y/h=0.5$ and the other half along $z/h=0.5$.  Note that half of the
information from the observations is redundant due to the diagonal symmetry of the flow.

\begin{figure}[!htbp]
\centering 
\includegraphics[width=0.5\textwidth]{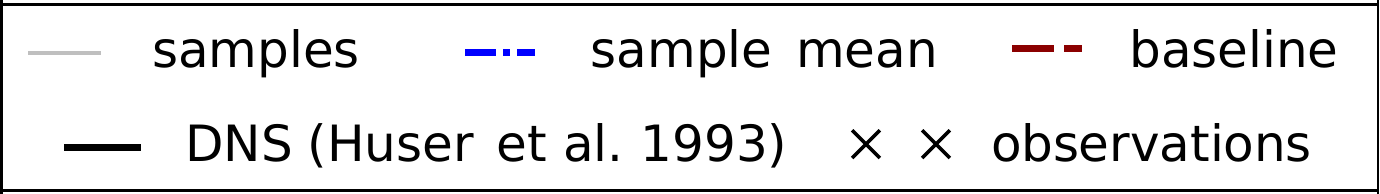}\\
\subfloat[Prior ensemble]{\includegraphics[width=0.45\textwidth]{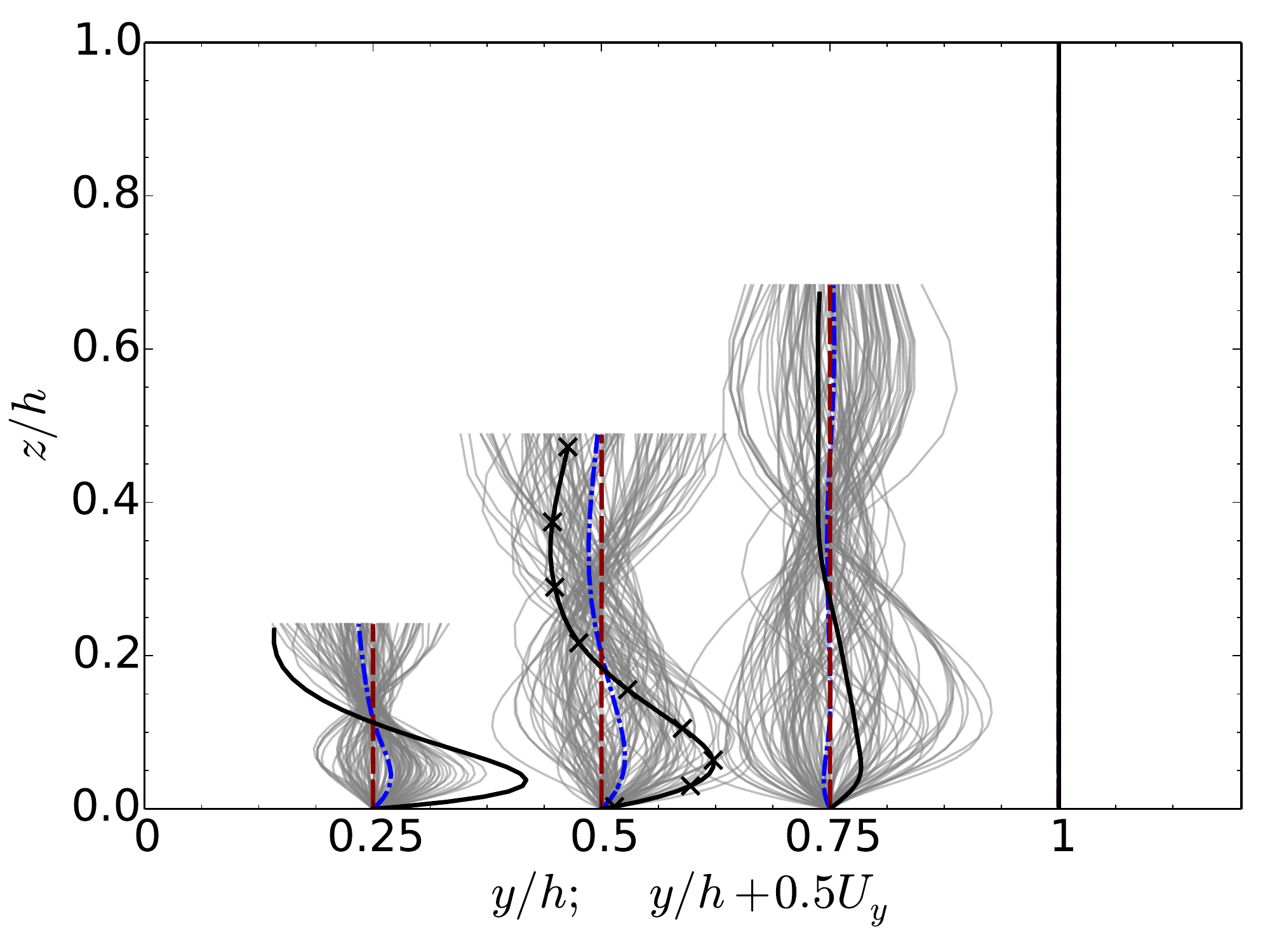}}\hspace{0.05em}
\subfloat[Posterior ensemble]{\includegraphics[width=0.45\textwidth]{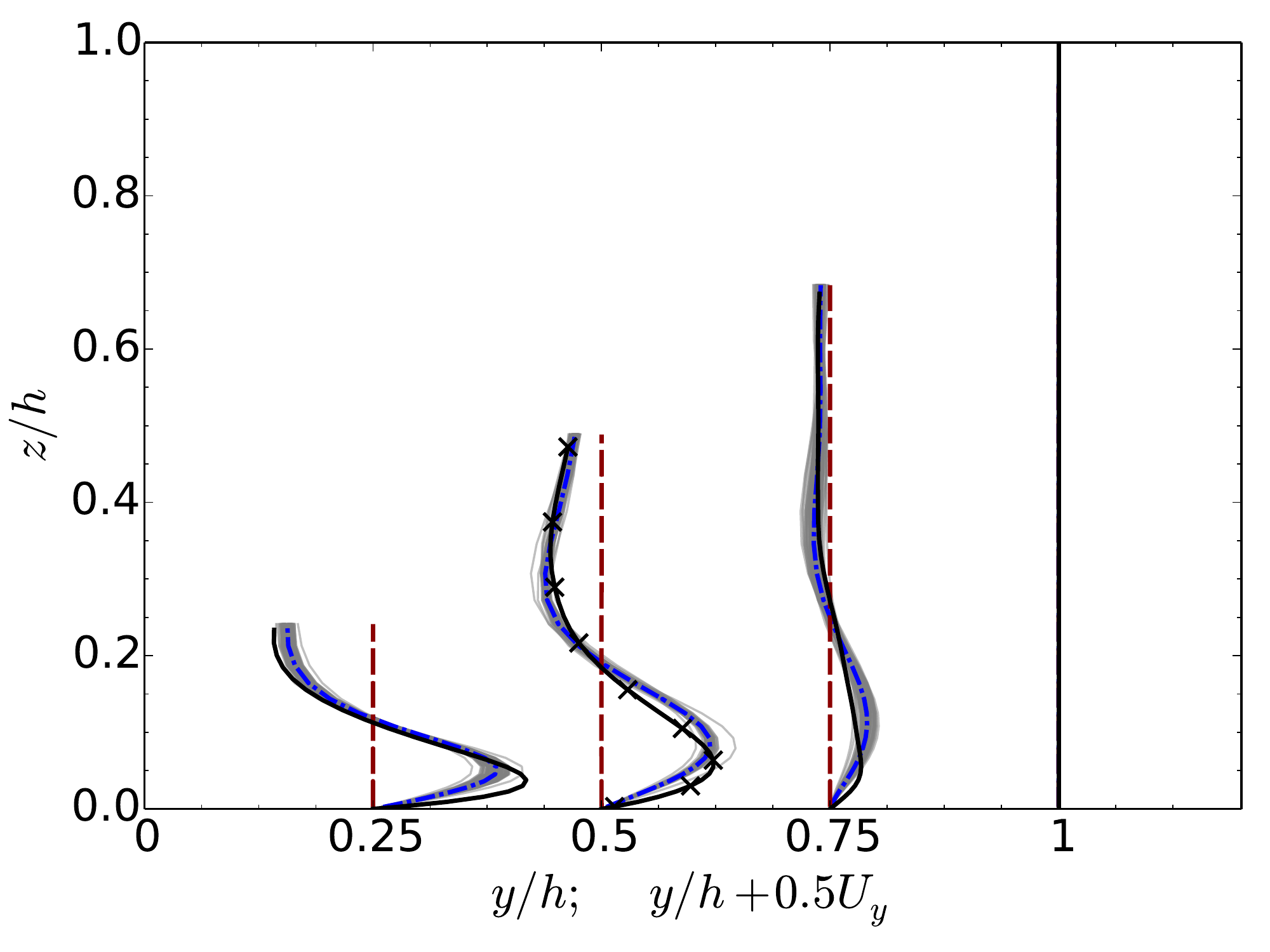}}\hspace{0.05em}
\caption{ (a) Prior velocity ensemble and (b) posterior velocity ensemble at four spanwise locations
  $y/h=0.25, 0.5, 0.75$ and $1$ with comparison to baseline and benchmark results. The velocity
  $U_z$ in the $z$ direction have similar characteristics and are thus omitted. The velocity
  profiles in the prior ensemble are scaled by a factor of 0.3 for clarity.}
\label{fig:U_duct}
\end{figure}

The ability of a numerical model to predict the secondary flow in $y$--$z$ plane is of most interest
for the flow in square duct. Therefore, the in-plane velocity field is identified as the QoI for
this case.  The in-plane flow velocity ($U_y$) on the four cross-sections as indicated in
Fig.~\ref{fig:domain_duct} are presented to facilitate quantitative comparison with the baseline and
benchmark results.  The velocity profiles $U_z$ in the $z$ direction have similar characteristics as
$U_y$ (but are not identical) and are thus omitted. The prior and posterior ensembles of the
velocity profiles for $U_y$ are shown in Fig.~\ref{fig:U_duct}.  Only the velocity profiles in the
region below the diagonal are presented due to the diagonal symmetry. It can be seen from
Fig.~\ref{fig:U_duct} that the baseline RANS simulation predicts uniformly zero in-plane velocities
as expected.  Around the baseline prediction, the prior ensembles are scattered due to the
perturbation of $\delta_{\xi}$ and $\delta_{\eta}$. The large range of scattering indicates that the
secondary flow is sensitive to the anisotropy of Reynolds stresses tensor, which has been reported
in previous studies~\cite{emory14estimate,huser1993direct}. Compared to the prior ensemble mean and
the baseline RANS prediction, the posterior ensemble mean of the velocities are significantly
improved along all four cross sections, as shown by good agreements with the benchmark data.  The
scattering has been significantly reduced as well, while still covering the truth adequately in most
regions.  The remaining differences and the regions where the ensemble fails to cover the truth can
be explained similarly as in the periodic hill case.  Similar to that in the periodic hill case,

\begin{figure}[!htbp]
\centering
\includegraphics[width=0.8\textwidth]{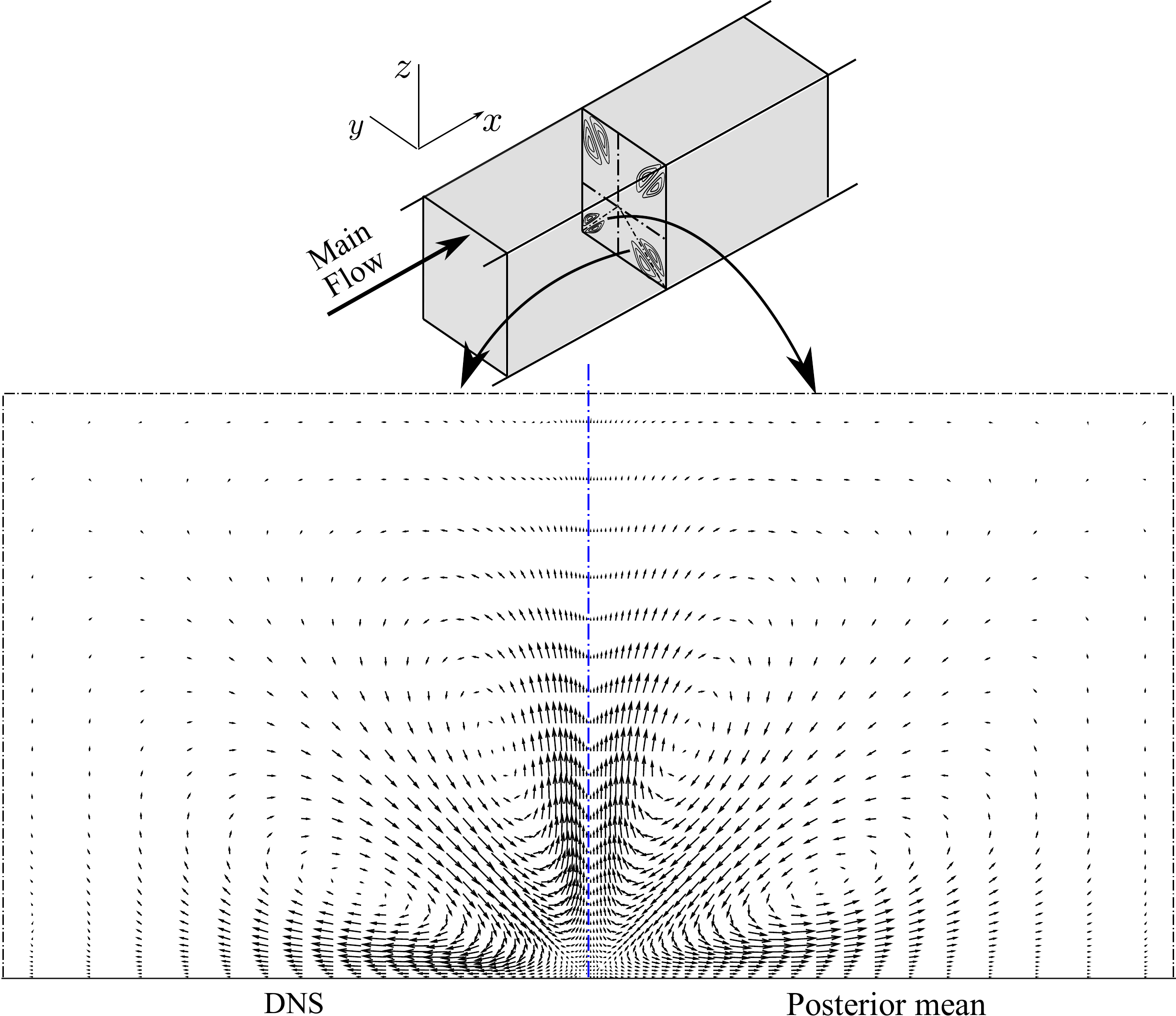}
\caption{Comparison of the velocity field in a square duct between the posterior mean and benchmark
  DNS data.The length and direction of an arrow indicate the magnitude and direction, respectively,
  of the in-plane flow velocity.  The plots are arranged such that a perfect agreement between the
  two would show as exact symmetry of the two panels along the vertical center line.  The vector
  field from the baseline RANS prediction is omitted since it is uniformly zero. }
\label{fig:U_vector_duct}
\end{figure}

Figure~\ref{fig:U_vector_duct} shows a comparison of the posterior ensemble mean field of the
in-plane flow velocity and the benchmark data. They are presented as vector plots to show the
overall features of secondary flow and particularly the vortex structure. The length and
direction of an arrow indicate the magnitude and direction, respectively, of the in-plane flow
velocity at that location.  The plots are arranged such that a perfect agreement between the two
would show as exact symmetry of the two panels along the center line.  The vector plot of the
velocity field from the baseline RANS prediction is omitted since it is uniformly zero. It can be
seen that the posterior ensemble mean demonstrates a very good agreement with the benchmark data in
most aspects, i.e., the direction and the intensity of secondary flow at most locations as well as
the center of vortex structure. Only minor differences between the two can be identified. For
example, the posterior mean velocity has a slightly smaller gradient of velocity magnitude compared
to the benchmark results near the symmetry line. The agreement clearly demonstrates the merits of
the current framework, particularly considering the fact that most of the commonly used turbulence
models are not capable of predicting the in-plane flow.  Specifically, all isotropic eddy viscosity
models completely miss the secondary flow, which is explained by the negligible
$\partial{U_y}/\partial{y}$ and $\partial{U_z}/\partial{z}$ terms and the Boussinesq assumption that
Reynolds stress is proportional to local strain rate of the mean flow. Even advanced models (e.g.,
Reynolds stress transport models) tend to underestimate the flow
intensity~\cite{demuren84}. Admittedly, velocity observations at some locations are used in this
method, but the amount of data used in the inference is rather small compared to the total degrees of
freedom of the Reynolds stress field.

\begin{figure}[!htbp]
  \centering
  \includegraphics[width=0.8\textwidth]{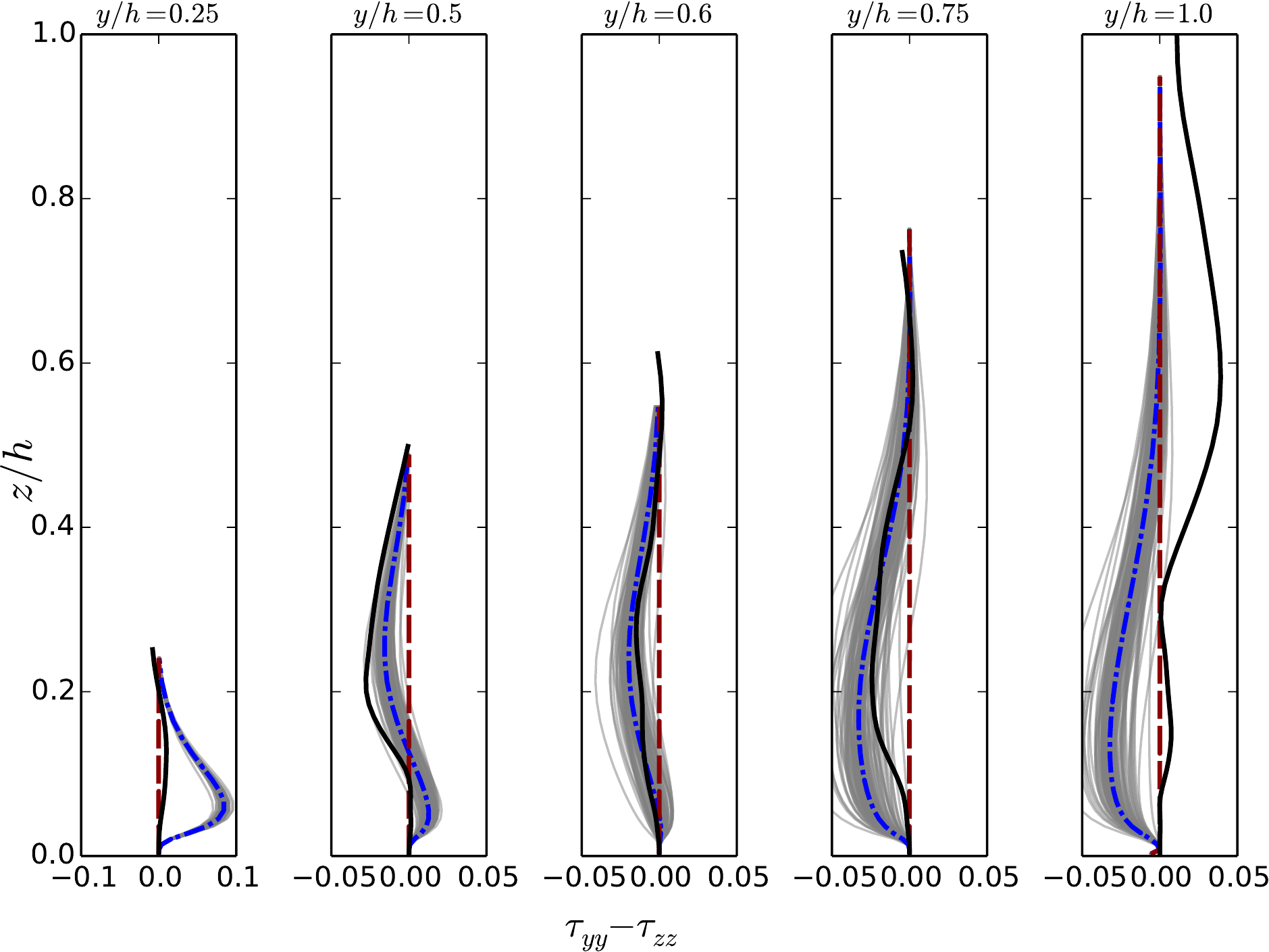}
  \caption{Comparison of the normal stresses imbalance at five locations $y/h = 0.25, 0.5, 0.6, 0.75$
    and $1.0$. A larger horizontal axis range is used in the panel for $y/h = 0.25$ due the large
    range of values $\tau_{yy}-\tau_{zz}$ at this location.}
  \label{fig:vw_duct}
\end{figure}

As mentioned above, the normal stress imbalance $\tau_{yy}-\tau_{zz}$ is the main driving force of
the secondary flow.  Therefore, the prior and posterior ensembles of the imbalance at five
locations, $y/h = 0.25, 0.5, 0.6, 0.75$, and $1$, are presented in Fig.~\ref{fig:vw_duct}.  It can
be seen that the baseline RANS prediction of $\tau_{yy}-\tau_{zz}$ is zero. Compared to the baseline
RANS prediction, the posterior normal stress imbalance shows a significant improvement in regions
close to the observations ($y/h=0.5$), although differences still exist, especially in the regions
far away from the observations (e.g., at $y/h = 1$). It is consistent with the argument made in
Section~\ref{sec:pripr-prob-spec} that the correlation decreases with distance, and that the quality
of correction heavily depends on correlations. Note that the inferred stress imbalances at
$y/h=0.75$ agree with the benchmark much better than do those at $y/h=0.25$,
although they have approximately the same distances from the observations, which are
distributed along $y/h = 0.5$. This can be explained by the fact that the length scale of the flow
decreases towards the corner (e.g., near $y/h=0.25$) due to the constriction of the duct walls, and
thus the correlation decreases much faster in this region than near the symmetry line.

\section{Discussion}
\label{sec:discuss}

\subsection{Computational cost of the model-form uncertainty quantification}
\label{sec:cost}
As mentioned in Sections~\ref{sec:implement}.  The ensemble Kalman method uses 60 samples (see
Table~\ref{tab:paraDA}) and needs approximately 10 iterations to achieve statistical
convergence. Therefore, each uncertainty quantification case involves 600 evaluations of the forward
RANS model \texttt{tauFoam}. Since each forward RANS evaluation is only 10\% as expensive as a
baseline RANS simulation (see Section~\ref{sec:implement}), the total computational cost of the
uncertainty quantification procedure is 60 times as that of the baseline simulation. However, note
that the propagation of samples can be done in parallel, i.e., in each iteration the 60 forward RANS
simulations were run simultaneously on 60 CPU cores. As a result, the wall time of the uncertainty
quantification procedure is approximately the same as that of the baseline simulation, assuming the
latter is run on a single core. Finally, the computational costs associated with the projection of
Reynolds stresses, the KL expansion, and the Kalman filtering are all neglected in the analysis
above. This is justified because the computational cost of the uncertainty quantification is indeed
dominated by the forward model evaluations.

\subsection{The role of correlation in current framework} 
\label{sec:correlation}
It has been pointed out that the relatively poor inference performance is expected in the vicinity of the
  crest, which is due to the lack of observations in the region and statistically significant
  correlations with the regions that have observations. The concept of correlation plays an important
  role in the current inference framework and warrants further discussions.

In three-dimensional complex flows more measurement data may be needed to obtain results of
  similar quality as presented here. In particular, when the flow consists of a number of distinct
  regions that are weakly correlated, a design of experiment study is needed to ensure all 
  regions of interest have measurements data if possible.
  However, note that the correlation within the flow field can be studied a priori solely based on
  an ensemble of RANS simulations before any measurements are performed. Based on the study of
  correlations, the measurements can subsequently be optimized. Therefore, such simulation-informed
  experimental design is feasible in practice.  We have performed such a correlation study in a more
  complex flow, the flow over a wing-body junction, and the detailed results are presented in
  ref.~\cite{wu2016wing}.

It is essential to choose proper correlation length scales based on that of the mean flow, which is
part of the physics-based prior knowledge. While an overly small length scale would fail to make
corrections to the regions without observation, an overly large length scale would lead to spurious
corrections. The correlation structure (e.g., of the velocities) in a flow field is very complex and
difficult to visualize due to the high dimensionality of the state. However, we can illustrate the
idea by considering the streamlines of the mean flow.  Intuitively, velocities at two points on the
same streamline should have a relatively high correlation. Consequently, observing the velocity at
one point can inform us about velocities at other points on the same streamline. Two points in
different coherent structures or regions as mentioned above, e.g., one point in the recirculation
zone and another in the shear zone, are likely to be on different streamlines. This explains why
points within the same region have higher correlations than the correlations among different
regions. That also justifies the arrangement of observation points shown in
Fig.~\ref{fig:domain_pehill} with observations scattered in all three regions of interest.  This
explanation of correlation structure is of course a highly simplified picture. In reality, fluid
flows are highly complex, coupled dynamic systems.  Velocities at different points can be correlated
due to continuity requirements and pressure. It is well known that the pressure is described by an
elliptic equation (for incompressible flows), which has whole-domain coupling characteristics.

\subsection{Success and limitation of the current framework}
\label{sec:success}
The overall idea of the proposed method is to improve flow field and QoI predictions and to quantify the
uncertainties therein by combining all sources of available information, including observation data,
physical prior knowledge, and RANS model predictions.  A Bayesian framework based on an iterative
ensemble Kalman method is used for the uncertainty quantification.  Numerical simulation results
have demonstrated the feasibility of the framework.  In particular, even with velocity observations
at very few locations, the posterior velocities are significantly improved compared to the baseline
results.

One may also expect that the uncertainties in the modeled Reynolds stresses can be quantified and
reduced. Indeed, the posterior ensemble obtained from the Bayesian inference process also has
information on the Reynolds stresses. However, our experience suggests that the posterior mean of an
arbitrarily chosen component or projection of the Reynolds stresses is not significantly more
accurate than those of the baseline prediction.

This apparent contradiction can be explained from two perspectives: the high dimensionality of the
Reynolds stress field and the mapping $\boldsymbol{\tau} \mapsto \mathbf{u}$ from Reynolds stresses
to mean velocities.  The most straightforward reason as mentioned in Section~\ref{sec:enkf} is that
the Reynolds stress discrepancy is a tensor field in a high-dimensional uncertainty space, and thus
the amount of velocity data is not sufficient to constrain its uncertainties, even when other prior
information is considered.  Moreover, the RANS equations describe a many-to-one mapping from
Reynolds stresses to mean velocities, and thus the mapping $\boldsymbol{\tau} \mapsto \mathbf{u}$ is
not invertible (i.e., a given velocity field may correspond to many possible Reynolds stress
fields). This is evident from the fact that the divergence of the Reynolds stress tensor, rather
than the Reynolds stress itself, appears in the RANS equation. Although difficult to prove
rigorously, we postulate that even Reynolds stress fields that have different divergences can map to
very similar velocity fields.  One can loosely think of the velocity field as being driven by a
\emph{projection} of the Reynolds stress on a \emph{low-dimensional manifold}.  The specific form of
the projection depends on the physics of individual flow. Taking the flow in a square duct for
example, it has been demonstrated by analytical
derivations~\cite{perkins1970formation,huser1993direct, emory14estimate} that the secondary flow is
primarily generated by the normal stress imbalance field $\tau_{yy} - \tau_{zz}$, or more precisely,
its cross spatial derivative $\frac{\partial^2}{\partial y \partial z} (\tau_{yy} - \tau_{zz})$. The
imbalance scalar can be obtained from the Reynolds stress tensor through linear mapping described by
a rank deficient matrix.  Since only velocity observation data are used in the Bayesian inference in
this work, one can only reasonably expect to infer the projection (i.e., the imbalance), but not the
full Reynolds stress field. This can be partly explained by the fact that the projection has a lower
dimension, but more importantly, the projection is an observable variable from a control-theoretic
perspective. This has been demonstrated in Fig.~\ref{fig:vw_duct}. The mapping between Reynolds
stresses and velocity as describe by the RANS equations are extremely complex due to the their
nonlinearity. This complexity and its implications to the current framework will be further
investigated.

Another limitation of the current framework lies in the iterative ensemble method used for the
uncertainty quantification, which is a computationally affordable method for approximate Bayesian
inference. The posterior distribution obtained with this method may deviate from the true
distribution.  This compromise is made in this work in consideration of the high computational costs
of RANS models (e.g., hours to days for realistic flow simulations), which makes more accurate
sampling methods such as those based on the Markov Chain Monte-Carlo method prohibitively expensive. The
accuracy of the ensemble-based method will be assessed in future work by comparing current results
with those obtained with MCMC, possibly by utilizing recently developed dimension-reduction methods
(e.g., active subspace methods~\cite{constantine2014active}, likelihood informed
dimension-reduction~\cite{cui2014likelihood}) and sampling techniques (e.g., delayed rejection
adaptive metropolis~\cite{haario2006dram}), or by building surrogate models to facilitate the MCMC
sampling.

\subsection{What if there are no observation data available?}
In light of the limitations of the framework as described above, two legitimate follow-up questions
can be raised. That is, given that the full Reynolds stress discrepancy field cannot be inferred
accurately from the velocity observations, (1) what would the value of the framework be in
engineering practice and (2) how can this framework can be used in scenarios with no observation
data.

Regarding the first question, sparse observation data are often available for engineering systems
that are in operation. For example, real-time monitoring sensors are often installed in wind farms,
nuclear power plants, and many other important facilities and devices. For these cases, the current
framework can provide a powerful method for combining information from the numerical models (often
greatly simplified due to stringent positive lead-time requirement in predictions), observation
data, and physical prior knowledge.

In scenarios where there are no observation data available as posed in the second question, the
current framework can be used in two ways. First, with the absence of observation data the inference
procedure essentially degenerates to forward uncertainty propagation, i.e., propagating the
uncertainties in the form of physical prior knowledge on Reynolds stresses to uncertainties in QoIs
(e.g., velocity, wall shear stresses, and reattachment point). This is somewhat similar to but more
comprehensive than the framework of Iaccarino and
co-workers~\cite{gorle2013framework,gorle2014deviation,emory2013modeling,emory2011modeling,emory14estimate},
since the prior in our work covers an uncertainty space rather than only a few limit states. Second,
when observation data are available in a geometrically similar case but perhaps at a lower Reynolds
number (e.g., the downscaled model in a laboratory experiment), the model uncertainties can first be
quantified and reduced with the data available on the scaled model. After the calibration, the
posterior Reynolds stress uncertainty distribution is extrapolated to the case of concern (e.g., the
flow in a geometrically similar prototype at a higher Reynolds number) to make predictions. Dow and
Wang~\cite{dow11quanti} used a similar calibration--prediction procedure to predict flows in
channels of different geometries by using Gaussian processes describing the eddy viscosity
discrepancy. Similar ideas have been suggested and advocated by Duraisamy et
al.~\cite{tracey2015machine,parish2016paradigm}. In all cases the calibration--prediction procedure
relies upon a crucial assumption that the calibration case and the prediction case share physically
similar characteristics, despite the differences in specific flow conditions (e.g., Reynolds number
or geometry). The feasibility of the calibration--prediction method based on the current framework
has been preliminarily explored by Wu et al.~\cite{wu2016bayesian}, which showed promising results
when the calibrated Reynolds stress discrepancies are used to predict flows in the same geometry but
at a Reynolds number one order of magnitude higher.  Prediction of flows in a different geometry, on
the other hand, has achieved less successes.  However, extreme caution must be exercised and expert
opinions must be consulted when using such an extrapolation method as presented in
\cite{wu2016bayesian}, since even a slight change of Reynolds number can lead to significant changes
of flow characteristics.  Ultimately, the use of this assumption has to be the judgment of the user,
which is clearly undesirable.  An improved, more intelligent framework should be sought for.  It
seems that modern machine learning methods have the potential of alleviating users of such burdens,
which is a topic of current research~\cite{tracey2015machine}.

\section{Conclusion}
\label{sec:conclude}

In this work we propose an open-box, physics-informed, Bayesian framework for quantifying and
reducing model-form uncertainties in RANS simulations. Uncertainties are introduced directly to the
Reynolds stresses and are represented with compact parameterization accounting for empirical prior
knowledge and physical constraints (e.g., realizability, smoothness, and symmetry).  An iterative
ensemble Kalman method is used to incorporate the prior knowledge with available observation data in
a Bayesian framework and propagate the uncertainties to posterior distributions of the Reynolds
stresses and other QoIs. Two test cases, the flow over periodic hills
and the flow in a square duct, have been used to demonstrate the feasibility and to evaluate the
performance of the proposed framework. Simulation results suggest that even with sparse
observations, the obtained posterior mean velocities have significantly better agreement with the
benchmark data compared to the baseline results. The methodology provides a general framework for
combining information from physical prior knowledge, observation data, and low-fidelity numerical
models (including RANS models and beyond) that are frequently used in engineering practice.

A notable limitation is that the full Reynolds stress field inferred from this method is not
accurate. This is attributed to the high dimension of the Reynolds stress uncertainty space, the
sparseness of the velocity observation data, and the nonlinear, possibly even non-unique, mapping
between the Reynolds stresses and velocities as described by the RANS equations. However, we argue
that the inferred Reynolds stresses are still valuable despite this limitation, and that they can be
extrapolated to cases with similar physical characteristics.  Another limitation of the current
framework lies in the iterative ensemble method used for the uncertainty quantification, which is
computationally less intensive but less accurate than exact Bayesian inference based on Markov Chain
Monte-Carlo sampling. The impact of the approximate Bayesian inference method will be investigated
in future studies.

\section{Acknowledgment}
We thank the anonymous reviewers for their comments, which helped in improving the quality and
clarity of the manuscript.  We gratefully acknowledge partial funding of graduate research
assistantships for JLW, JXW, and RS from the Institute for Critical Technology and Applied Science
(ICTAS, Grant number 175258).





 \appendix

\section{Mapping from Barycentric Coordinates to Natural Coordinates}
\label{app:mapping}

Following the work of Iaccarino et al., we introduce uncertainties (also referred to as
perturbations) to the Reynolds stresses by perturbing its magnitude (the turbulent kinetic energy
$k$) and the shape (the eigenvalues $\lambda_1$ and $\lambda_2$ of the anisotropy tensor) as shown
in Eq.~(\ref{eq:tau-decomp}).  The eigenvalues can be linearly transformed to the Barycentric
coordinate $(C_1, C_2, C_3)$ as follows~\cite{banerjee2007presentation,emory2011modeling}:
\begin{subequations}
  \label{eq:lambda2c}
\begin{align}
  C_1 & = \lambda_1 - \lambda_2 \\
  C_2 & = 2(\lambda_2 - \lambda_3) \\
C_3 & = 3 \lambda_3 + 1
\end{align}  
\end{subequations}
where $C_1$, $C_2$, and $C_3$ indicate the portion of areas of the three sub-triangles in the
Barycentric triangle, and thus they sum to 1.  Placing the triangle in a Cartesian coordinate
$\mathbf{x}^b \equiv (y^b, y^b)$, the location of any point within the triangle is a convex
combination of those of the three vertices, i.e., 
\begin{equation}
  \label{eq:c2alpha}
\mathbf{x}^b =  \mathbf{x}^b_{1c} C_1 + \mathbf{x}^b_{2c} C_2  + \mathbf{x}^b_{3c} C_3
\end{equation}
where $\mathbf{x}^b_{1c}$, $\mathbf{x}^b_{2c}$, and $\mathbf{x}^b_{3c}$ are the coordinates of the
three vertices of the triangle (see Fig.~\ref{fig:bary}). The superscript $b$ is used to distinguish
it from the coordinate system for the fluid flow problems.

While Emory\cite{emory2011modeling} perturbed the Reynolds stress towards the three limiting
states (the vertices of the triangle), we need to parameterize and explore the entire triangle. To
facilitate parameterization with minimum artificial capping of Reynolds stresses falling outside the
realizable range, we further transform the Cartesian coordinate $(x^b, y^b)$ to the natural
coordinate $(\xi, \eta)$ by using the standard finite element shape functions:
\begin{subequations}
\label{eq:xi2x}
\begin{align}
  x^b & = x(\xi, \eta) = \sum_{i=1}^4 N_i (\xi, \eta) x^b_i \\
  y^b & = y(\xi, \eta) = \sum_{i=1}^4 N_i (\xi, \eta) y^b_i
\end{align}
\end{subequations}
where $(x^b_i, y^b_i)$ are the coordinates of four vertices, and $N_1$, $N_2$, $N_3$, and $N_4$ are
shape functions defined as
\begin{subequations}
  \label{eq:shapefun}
\begin{align}
  N_1(\xi, \eta) & = \frac{(1 - \xi)(1 - \eta)}{4}  \notag \\
  N_2(\xi, \eta) & = \frac{(1 + \xi)(1 - \eta)}{4}  \notag \\
  N_3(\xi, \eta) & = \frac{(1 + \xi) (1 + \eta)}{4}  \notag \\
  N_4(\xi, \eta) & = \frac{(1 - \xi) (1 + \eta)}{4}. \notag
\end{align}
\end{subequations}

The mapping from the natural coordinate $(\xi, \eta)$ to the physical coordinate $(x^b, y^b)$ as shown in
Eq.~(\ref{eq:xi2x}) is routinely used in finite element methods.  However, the inverse mapping,
i.e., computing the natural coordinate $(\xi, \eta)$ for a given physical coordinate $(x^b, y^b)$, is
nontrivial and uncommon due to the difficulty of solving the bilinear equation system
Eq.~(\ref{eq:xi2x}). In this work we use the analytical results from~\cite{hua1990ai} to obtain this
mapping.

In summary, the Reynolds stresses field $\bstaurans$ computed from the baseline RANS
simulation are mapped to the physical interpretable variables $\tilde{k}^{rans},
\tilde{\xi}^{rans}, \tilde{\eta}^{rans}$ via the following sequence:
\begin{equation}
\label{eq:forward-mappings}
\tilde{\bs{\tau}}
\xrightarrow{(\ref{eq:tau-decomp})} (\tilde{k}, \tilde{\lambda}_1, \tilde{\lambda}_2)
\xrightarrow{(\ref{eq:lambda2c})} (\tilde{k}, \tilde{C}_1, \tilde{C}_2)
\xrightarrow{(\ref{eq:c2alpha})} (\tilde{k}, \tilde{x}^{b}, \tilde{y}^{b})
\xrightarrow{\textrm{inv.~of } (\ref{eq:xi2x})} (\tilde{k}, \tilde{\xi}, \tilde{\eta})
\notag
\end{equation}
where unperturbed quantities $\mathbf{v}^{rans}_1$, $\mathbf{v}^{rans}_2$, and
$\mathbf{v}^{rans}_3$, dependent variables $\lambda_3$ and $C_3$, and superscript $rans$ are omitted
for simplicity of notation.  Equations describing the mappings are indicated above the corresponding
arrow. Equation~(\ref{eq:tau-decomp}) indicates eigen-decomposition and reconstruction. After the
sequence of mapping, uncertainties are introduced into these transformed quantities by modeling the
truth of $k$, $\xi$, $\eta$ as random fields with their respective baseline results as priors (see
Eq.~(\ref{eq:delta-def})). They are subsequently used to obtain Reynolds stresses via the inverse of
mapping sequence as above:
\begin{equation}
\label{eq:back-mappings}
 ({k}, {\xi}, {\eta}) \xrightarrow{(\ref{eq:xi2x})}
 ({k}, {x}^{b}, {y}^{b}) \xrightarrow{\textrm{inv.~of } (\ref{eq:c2alpha})}
({k}, {C}_1, {C}_2) \xrightarrow{\textrm{inv.~of } (\ref{eq:lambda2c})} 
({k}, {\lambda}_1, {\lambda}_2) \xrightarrow{(\ref{eq:tau-decomp})} 
\bs{\tau}
\notag
\end{equation}

\section{Iterative Ensemble Kalman Method for Inverse Modeling}
\label{app:enkf}

The algorithm of the iterative ensemble Kalman method for inverse modeling is summarized
below. See~\cite{iglesias2013ensemble} for details.

Given velocity prediction from the baseline RANS simulation $\mathbf{u}^{rans}$ and observations with error
covariance matrix $R$, the following steps are performed:
\begin{enumerate}
\item \textbf{(Sampling step)} Generate initial ensemble $\{{\mathbf{x}_j}\}_{j = 1}^N$ of size $N$,
  where the augmented system state is:
  \begin{equation}
    \label{eq:ini-x}
    \mathbf{x}_j  = [\mathbf{u}^{rans}, \bs{\omega}]_j  
    \notag
  \end{equation}

\item \textbf{(Prediction step)} 
  \begin{enumerate}
  \item Propagate the state from current state $n$ to the next iteration level $n+1$ with the forward model
    \texttt{tauFoam}, indicated as $\mathcal{F}$,
  \begin{equation}
    \label{eq:forward}
    \hat{\mathbf{x}}_j^{(n+1)} = \mathcal{F} [ \mathbf{x}_j^{(n)} ]
    \notag
  \end{equation}
  This step involves reconstructing Reynolds stress fields for each sample and computing the
  velocities from the RANS equations.
\item
  Estimate the
  mean $\bar{\mathbf{x}}$ and covariance $P^{(n+1)}$ of the ensemble as:
  \begin{subequations}
    \begin{align}
      \bar{\mathbf{x}}^{(n+1)} = \frac{1}{N}\sum_{j=1}^N{\hat{\mathbf{x}}^{(n+1)}_j}     \notag  \\ 
      P^{(n+1)} = \frac{1}{N-1} \sum_{j = 1}^{N} {\left( \hat{\mathbf{x}}_j\hat{\mathbf{x}}_j^T -  
          \bar{\mathbf{x}}\bar{\mathbf{x}}^T \right)^{(n+1)}} \notag   
    \end{align}
  \end{subequations}
\end{enumerate}

\item  \textbf{(Analysis step)}
  \begin{enumerate}
  \item Compute the Kalman gain matrix as:
    \begin{equation}
      \label{eq:kalman-gain}
      K^{(n+1)} = P^{(n+1)} H^T (H P^{(n+1)} H^T + R)^{-1} 
      \notag
    \end{equation}

  \item Update each sample in the predicted ensemble as follows:
        \begin{equation}
      \label{eq:update}
      \mathbf{x}_j^{(n+1)} = \hat{\mathbf{x}}_j^{(n+1)} + K (\mathbf{y} - H
      \hat{\mathbf{x}}_j^{(n+1)})  
      \notag
    \end{equation}
  \end{enumerate}
  The vector $\mathbf{y}$ represents observation and $H$ is the observation matrix, which maps state
  space to the observation space.

\item Repeat the prediction and analysis steps until  the ensemble is statistically converged.
\end{enumerate}





\end{document}